\documentclass[usenatbib]{mnras}

\usepackage{txfonts}

\usepackage[T1]{fontenc}
\usepackage{ae,aecompl}

\usepackage{graphicx}	
\usepackage{amssymb}	
\usepackage{changes}
\usepackage{relsize}    

\def\apj{ApJ}
\def\apjl{ApJL}
\def\apjs{ApJ Suppl. Ser.}

\def\aap{A\&A}

\def\mnras{MNRAS}

\def\pasj{PASJ}

\def\beq#1{\begin{equation}\label{#1}}
\def\eeq{\end{equation}}
\def\beqa#1{\begin{eqnarray}\label{#1}}
\def\eeqa{\end{eqnarray}}

\def\Eq#1{Eq.~(\ref{#1})} 
\def\eqn#1{~(\ref{#1})}
\def\myfrac#1#2{\left(\frac{#1}{#2}\right)}

\def\apgt{\ {\raise-.5ex\hbox{$\buildrel>\over\sim$}}\ }
\def\aplt{\ {\raise-.5ex\hbox{$\buildrel<\over\sim$}}\ }
\newcommand{\ms}{$M_\odot$}
\newcommand{\msun}{$M_\odot$}
\newcommand{\kms}{\:\mbox{km/s}}
\newcommand{\porb}{\mbox {$P_{\mathrm{ orb}}$}}
\newcommand{\mch}{$\mathrm{M_{Ch}}$}
\newcommand{\myr}{\mbox {~${\rm M_{\odot}~yr^{-1}}$}}

\title[Symbiotic X-ray binaries]{Wind-accreting Symbiotic X-ray Binaries}

\author[L.R. Yungelson et al.]{
Lev R. Yungelson,$^{1}$\thanks{E-mail: lev.yungelson@gmail.com)}
Alexandre G. Kuranov$^{2,3}$
and Konstantin A. Postnov$^{2,4}$
\\
$^{1}$ Institute of Astronomy, Russian Academy of Sciences, Pyatnitskaya 48, 119017 Moscow, Russia\\
$^{2}$ Sternberg Astronomical Institute, M.V. Lomonosov Moscow State University, 13, Universitetskij pr., 119234 Moscow, Russia\\
$^{3}$ Russian Foreign Trade Academy,  4a Pudovkin str., 119285 Moscow, Russia\\
$^{4}$ Kazan Federal University, Kremlevskaya 18, 420008 Kazan, Russia
}
\date{Accepted XXX. Received YYY; in original form ZZZ}

\pubyear{2018}

\begin{document} 
\label{firstpage}
\pagerange{\pageref{firstpage}--\pageref{lastpage}}
\maketitle

\begin{abstract}
We present a new model of the population of symbiotic X-ray binaries (SyXBs)
that takes into account non-stationary character of quasi-spherical sub-sonic
accretion of the red giant's stellar wind onto slowly rotating neutron stars.
Updates of the earlier models are given, which include more strict criteria of
slow NS rotation for plasma entry into the NS magnetosphere via Rayleigh-Taylor
instability, as well as more strict conditions for settling accretion for slow
stellar winds, with an account of variations in the specific angular momentum of
captured stellar wind in eccentric binaries. These modifications enabled a more
adequate description of the distributions of observed systems over binary orbital
periods, NS spin periods and their X-ray luminosity in the $\sim
10^{32}-10^{36}$~erg s$^{-1}$ range and brought their model Galactic number into reasonable agreement with the observed one. Reconciliation of the model and
observed orbital periods of SyXBs requires a low efficiency of matter expulsion
from common envelopes during the evolution that results in the formation of
NS-components of symbiotic X-ray systems. 

\end{abstract}




\section{Introduction}
\label{sec:intro}

Symbiotic X-ray binaries (SyXBs) represent a small group of binary systems
hosting an accreting neutron star (NS) and a late-type giant (K1-M8). Candidate
systems CGCS~5926 \citep{2011A&A...534A..89M} and CXOGBS J173620.2-29333
\citep{2014ApJ...780...11H} possibly host carbon stars. Despite the fact that
the first SyXB was identified more than 40~yr ago \citep{1977ApJ...211..866D},
currently, only about a dozen of SyXBs are detected (see Table \ref{t:syxbs}
listing the basic parameters of objects known at the time of writing).

\begin{table*}
  \caption{Parameters of observed and suspected SyXBs. Listed are the source name, the NS spin period $P^*$ (s), 
the binary orbital period $P_{\rm orb}$, 
the X-ray luminosity $L_{\rm  X}$, the source distance  $d$. 
}
\label{t:syxbs}
\begin{tabular}{lcccccc}
\hline
\hline \\[0.1mm]
 SyXB&$P^*$  &$P_{\rm orb}$ &$L_{\rm X}$ & d 
\\
            & (s)  & (day)       & (erg s$^{-1}$) & (Kpc) 
\\
\hline \\[0.05mm]
 GX 1+4 &$\simeq140$$^{(1)}$ &1161$^{(2, 3)}$&$10^{35}$---$10^{36}$$^{(4)}$&4.3$^{(2)}$
\\
                     & & 295$\pm$70$^{(5)}$ & & &\\
                     & & 304$^{(6, 7)}$ & & &\\
4U 1954+319 & $\sim 18300$$^{(8)}$ & $\gtrsim400$?$^{(9)}$ & $4\times10^{32}$$^{(9)} $ &1.7$^{(9)}$
\\
4U~1700+24 &? & 404$\pm$20$^{(6)}$ & $2\times10^{32}-10^{34}$$^{\rm(10)}$ & $0.42\pm0.4$$^{(10)}$
\\
                     & & 4391$^{(23)}$ & & &\\
Sct X-1&113$^{(11)}$  & ? & $2\times10^{34}$$^{(11)}$ & $\geq4^{(11)}$\\
IGR J16194-2810 &?&?&$\leq 7\times10^{34}$$^{\rm
(12)}$&$\leq3.7^{(12)}$\\
IGR J16358-4726 & 5850$^{(13)}$ & ? & $3\times10^{32}-2\times10^{36}$$^{(14)}$ & 5-6; 12-13$^{(15)}$ \\
CGCS 5926 & ? & $\sim151^{(16)}$ & $\leq3\times10^{32(16)}$ & $5.2^{(16)}$\\ 
CXOGBS J173620.2-293338 & ? &  ?& $\sim9\times10^{32}$$^{(17)}$ &?
 \\
XTE J1743-363 &? &? &? &$\sim5^{(18)}$  
\\
XMMU~J174445.5-295044 & ? & ? & $\gtrsim 4\times10^{34}$$^{(19)}$& $3.1^{+1.8}_{-1.1}$$^{(19)}$ \\
3XMM~J181923.7-170616IGR  & $407.9$$^{(20)}$ & ? &  $2.78\times10^{34}d_{10}^2$$^{(20)}$ & ?\\
IGR~J17329-2731 & $6680\pm3$$^{(21)}$& ? & ?& $2.7^{+3.4}_{-1.2}$$^{(21)}$ \\
IGR~J17197-3010 &                      &   & $\lesssim1.6\times10^{35}$$^{(22)}$ & 6.3-16.6$^{(22)}$\\
\hline
\end{tabular}
\vskip 0.05cm
\begin{flushleft}
1 -  \citet{2007A&A...462..995F},
2 - \citet{2006ApJ...641..479H},
3 - \citet{2017A&A...601A.105I},
4 - \citet{GonzalezGalan2011}, 
5 - \citet{2016pas..conf..133M},
6 - \citet{2002ApJ...580.1065G}, 
7 - \citet{1999ApJ...526L.105P},
8 - \citet{2008ApJ...675.1424C},
9 - \citet{2006A&A...453..295M}, 
10 - \citet{2002A&A...382..104M},  
11 - \citet{2007ApJ...661..437K}, 
12 - \citet{2007A&A...470..331M}, 
13 - \citet{2004ApJ...602L..45P}, 
14 - \citet{2007ApJ...657..994P}, 
15 - \citet{2005A&A...444..821L}, \\
16 - \citet{2011A&A...534A..89M}, 
17 - \citet{2017AAS...23031704H}, 
18 - \citet{2012MNRAS.422.2661S}, 
19 - \citet{2014MNRAS.441..640B}, 
20 - \citet{2017ApJ...847...44Q},  
21 - \citet{2018A&A...613A..22B}, 
22 - \citet{2012A&A...538A.123M}, 
23 - \citet{2018arXiv181208811H}. 
\end{flushleft}
\end{table*}
In SyXBs, a NS can accrete matter lost by the low-mass red giant companion as a
relatively slow stellar wind ($v_\mathrm{w}\sim$ several 10 km s$^{-1}$) and
during the Roche lobe overflow. SyXBs are characterized by the relatively low
mean X-ray luminosity $L_\mathrm{x}\sim 10^{32}-10^{36}$~erg s$^{-1}$ and
manifest outburst activity typical for wind-accreting NSs. The relative
faintness and non-stationarity of the sources make it difficult to detect and
analyse the NS rotation, therefore the NS spin periods $P^*$ have been measured
only for a handful of sources (see Table \ref{t:syxbs}). Spin periods range from several hundred to several thousand seconds, which is related to the
character of accretion onto magnetized NSs (see below). The estimated
orbital periods of SyXBs $P_\mathrm{b}$ may be as high as several years (e.g., in GX 1+4). Ellipticity of orbits is detected for GX 1+4 ($e$=0.10,
\citet{2006ApJ...641..479H}) and 4U~1700+24 (e=$0.26\pm0.15$,
\citet{2002ApJ...580.1065G}).

SyXB studies provide an independent probe of stellar winds from the red giant
components and the physics of wind accretion onto NSs. The observed
characteristics of SyXBs can be reasonably reproduced in the model of settling
accretion onto slowly rotating magnetized NS developed by
\citet{2012MNRAS.420..216S,2018ASSL..454..331S},
with an account of the NS spin evolution in
binary systems \citep{2009ARep...53..915L}.
The Galactic population of SyXBs was
simulated in our earlier papers \citep{2012MNRAS.424.2265L,2015AstL...41..114K},
in which the algorithm of the calculations and the model are presented in more
detail.

The present study is motivated by further development of the settling
accretion model \citep{2017arXiv170203393S,2018ASSL..454..331S} and availability of new
observations of SyXBs. The main modifications of the accretion theory can be
summarised as follows. 
1) The model of stellar winds from low-mass
post-main-sequence stars is updated. The stellar wind velocity $v_\mathrm{w}$ of
low-mass giants, which is crucial for the gravitational capture of the matter by a
NS in a binary system is now calculated according to the prescription suggested
in \cite{2006MNRAS.372.1389L}. 
2) The criteria and parameters of the subsonic
settling accretion are improved in accordance with the latest understanding of
this regime. 
3) An account of the transitions between the Compton
and radiative plasma cooling with changing X-ray luminosity and the
photoionization heating of the accreting plasma at low relative wind velocities
is included in the model. 
4) The reduction of NS equilibrium periods in
eccentric binaries is taken into account. These amendments enabled us to reach a
better description of the observed location of SyXBs in the $P^*-L_\mathrm{x}$
diagram and to improve the agreement of predicted and observed Galactic numbers
of SyXBs.

The paper is organised as follows. 
The improvements in the settling accretion theory are briefly described in \S~\ref{s:settling}.
In \S~\ref{sec:method} 
we describe the main modifications of the model. 
Results of the modeling are presented in \S~\ref{s:results}. 
Discussion and conclusion follow in \S~\ref{s:disc}. 
In Appendix, we present the treatment of the effect of the orbital eccentricity on the NS equilibrium period at the settling accretion stage.

\section{Refinement of the settling accretion theory}
\label{s:settling}

The settling accretion theory is designed to describe the interaction of
accreting plasma with magnetospheres of slowly rotating NSs. The key feature of
the model is calculation of the steady plasma entry rate into the
magnetosphere due to the Rayleigh-Taylor (the RT) instability regulated by plasma cooling (Compton or radiative). RT instability can be suppressed by fast
rotation of the magnetosphere \citep{1980ApJ...235.1016A}. In the present study
we will use the criterion from the latter paper:
\beq{e:fastP}
P^*>27 [\mathrm{s}] \dot M_{16}^{1/5}\mu_{30}^{33/35}(M_\mathrm{x}/M_\odot)^{-97/70}\,.
\eeq
At faster NS periods, the accretion regime at any X-ray luminosity will be
supersonic because of efficient plasma penetration into the magnetosphere via
Kelvin-Helmholtz instability \citep{1983ApJ...266..175B}. Here and below, 
$\dot M = 10^{16} \dot M_{16} \mathrm{g\, s}^{-1}$ is the accretion rate,
$\mu=10^{30}\mu_{30} \mathrm{G\,cm}^3$ is the NS magnetic moment.
 
 It was found \citep{1984ApJ...278..326E,2012MNRAS.420..216S} that the Compton
cooling of accreting plasma by X-ray photons generated near the NS surface
enables a free-fall (Bondi-type) supersonic accretion onto the NS magnetosphere
provided that the X-ray luminosity is above some critical value $L_{x}^\dag\simeq
4\times 10^{36}$ erg s$^{-1}$. X-ray luminosity $L_\mathrm{x}$ of an accreting
NS is related to the mass accretion rate $\dot M_\mathrm{x}$ as
$L_\mathrm{x}\approx 0.1 \dot M_\mathrm{x} c^2$ ($c$ is the speed of light), and
this critical $L_\mathrm{x}$ corresponds to a mass accretion rate onto NS 
$\dot M_\mathrm{x}\simeq 4\times 10^{16}$~g s$^{-1}$. 

At lower X-ray luminosities, accreting plasma above the magnetosphere remains hot enough to enable effective inflow into the magnetosphere
due to RT instability, and the plasma entry rate is controlled by the plasma cooling rate. This results in the formation of a hot, convective shell above the magnetosphere.
In this shell, the plasma gravitationally captured by the NS from the stellar wind of the companion (basically, at the Bondi-Hoyle-Lyttleton rate, $\dot M_\mathrm{B}$) subsonically  settles down towards the NS magnetosphere at a rate 
$\dot M_\mathrm{x}\approx f(u)\dot M_\mathrm{B}$, with the dimensionless factor $f(u)\lesssim 0.5$\footnote{In a radial accretion flow with an account of the Compton cooling, the value $f(u) \gtrsim 0.5$ also corresponds to the location of the sonic point  in the flow below the magnetospheric boundary $R_\mathrm{A}$ \citep{2012MNRAS.420..216S} enabling a free-fall plasma flow down to 
$R_\mathrm{A}$ and the formation of a shock above the magnetosphere 
\citep{1976ApJ...207..914A}.}, whose precise value depends on the plasma cooling regime (Compton or radiative). 
With good accuracy, this factor can be written as 
\begin{equation}
\label{e:f(u)}
f(u) \approx \left [ \frac{t_\mathrm{ff}}{t_\mathrm{cool}} \right ]^{1/3},
\end{equation}
where $t_\mathrm{ff}$ is the free-fall time at the magnetospheric boundary (the Alfv\'en radius, $R_\mathrm{A}$), $t_\mathrm{cool}$ is the characteristic plasma cooling time. Clearly, in the case $t_\mathrm{cool}\gg t_\mathrm{ff}$, this factor can be very small, leading to an effective decrease in the mass accretion rate onto the NS surface compared to the maximum available Bondi rate, $\dot M_\mathrm{B}$. 

In a circular binary system with  component separation $a$, the Bondi-Hoyle-Lyttleton accretion rate can be estimated as
\beq{e:dotMB}
\dot M_\mathrm{B}\approx \frac{1}{4}\dot M_\mathrm{o}\frac{v_\mathrm{rel}}{v_\mathrm{w}}\myfrac{R_\mathrm{B}}{a}^{2}\,,
\eeq
where $\dot M_\mathrm{o}$ is the mass-loss rate from the optical star, the gravitational Bondi radius is
\beq{e:RB}
R_\mathrm{B}=\delta\frac{2GM_\mathrm{x}}{v_\mathrm{rel}^2+c_\mathrm{s}^2}.
\eeq
Here $M_\mathrm{x}$ is the NS mass,  
$v_\mathrm{rel}^2=v_\mathrm{orb,x}^2+v_\mathrm{w}^2$ is the relative stellar wind velocity, $v_\mathrm{orb,x}$ is the NS orbital velocity, $c_\mathrm{s}$ is the sound velocity in the matter. The latter can be ignored in cold stellar winds of late-type stars. The estimate \eqn{e:dotMB} assumes a spherically-symmetric wind outflow from the optical star and is applicable for $R_\mathrm{B}\ll a$. The numerical factor $\delta\sim 1$ in \Eq{e:RB} takes into account the actual location of the bow shock in the stellar wind \citep{1971MNRAS.154..141H}. 
Some modern numerical calculations \citep[see, e.g.][]{2017ApJ...846..117L} suggest a smaller (up to an order of magnitude) mass accretion rates onto a gravitating mass from the stellar wind than given by the standard Bondi-Hoyle-Lyttleton formula, which can be reformulated in terms of smaller values of the parameter $\delta\simeq 0.3-0.5$.
Other studies \citep[e.g.,][]{2017MNRAS.468.3408D} claim that accretion rate can exceed the Bondi one due to gravitational focusing of the wind. Therefore, it should be born in mind that the numerical factor $\delta$ in \Eq{e:RB} can differ from unity by a factor of a few.

The Alfv\'en radius is defined from the pressure balance between the accreting plasma and the magnetic field at the magnetospheric boundary and depends on the actual mass accretion rate, $\dot M_\mathrm{x}$ (i.e., the observable X-ray luminosity $L_\mathrm{x}$), and the NS magnetic momentum, $\mu$:
\beq{}
R_\mathrm{A}\sim \myfrac{f(u)\mu^2}{\dot M_\mathrm{x}\sqrt{GM_\mathrm{x}}}^{2/7}\,.
\eeq
The factor $f(u)$ equals 
\beq{e:fuC}
f(u)_\mathrm{Comp}\approx 0.22 \zeta^{7/11} \dot M_{x,16}^{4/11}\mu_{30}^{-1/11}
\eeq
or
\beq{e:furad}
f(u)_\mathrm{rad}\approx 0.1 \zeta^{14/81} \dot M_{x,16}^{6/27}\mu_{30}^{2/27}
\eeq
for the Compton and radiative cooling, respectively \citep{2017arXiv170203393S,2018ASSL..454..331S}. Here $\zeta\lesssim 1$ is the numerical factor determining the characteristic scale of the growing RT mode (in units of the Alfv\'en radius $R_\mathrm{A}$).  

Thus, specifying the plasma cooling mechanism, we are able to estimate the expected reduction in the mass accretion rate at the settling accretion stage, 
$f(u)=F(\dot M_\mathrm{x},...)=F(f(u)\dot M_\mathrm{B},...)$, and, by solving for $f(u)$, to find the explicit expression for $\dot M_\mathrm{x}$ as a function of $\dot M_\mathrm{B}$ and other parameters.

Putting all things together, we are able to express the expected accretion rate onto the NS (which can be directly probed by the observed X-ray luminosity 
$L_\mathrm{x}$) 
using the Bondi capture rate $\dot M_\mathrm{B}$, which can be calculated 
from the known mass-loss rate of the optical companion $\dot M_\mathrm{o}$, stellar wind velocity $v_\mathrm{w}$ and binary system parameters:
\beq{e:MxC}
\dot M_{x,16}^\mathrm{Comp}\simeq 0.1 \zeta \dot M_{B,16}^{11/7}\mu_{30}^{-1/7}
\eeq
for the Compton cooling and 
\beq{e:Mxrad}
\dot M_{x,16}^\mathrm{rad}\simeq 0.05\zeta^{2/9}
\dot M_{B,16}^{9/7}\mu_{30}^{2/21}
\eeq
for the radiative cooling. The actual accretion rate onto the NS is taken to be 
$\dot M_\mathrm{x}=\max\{\dot M_{x}^\mathrm{Comp},\dot M_{x}^\mathrm{rad}$\} and determines the actual plasma cooling regime. Matching of \Eq{e:MxC} and \Eq{e:Mxrad} shows that with a given NS magnetic momentum, the Compton cooling of plasma dominates for the gravitational capture mass rate 
\beq{e:CvsRad}
\dot M_{B,16}\gtrsim 0.1 \zeta^{-2/9}\mu_{30}^{5/6}\,.
\eeq
As the parameter $\zeta$ is always less than 1, this explains why in our previous studies \citep{2015AstL...41..114K} the Compton cooling was ignored for low mass-accretion rates. However, it becomes important during outbursts caused by the stellar wind parameters variations (see below).

During the settling accretion, a hot convective shell formed above the NS magnetosphere mediates the angular momentum transfer to/from the rotating NS, enabling long-term spin-down episodes with spin-down torques correlated with the X-ray luminosity, as observed, for example, in GX~1+4 
\citep{1997ApJ...481L.101C,GonzalezGalan2011}. 
Turbulent stresses acting in the shell lead to an almost iso-momentum angular velocity radial distribution, $\omega(r)\sim 1/r^2$, suggesting the conservation of 
the specific angular momentum of gas captured near the Bondi radius 
$R_\mathrm{B}$, $j_\mathrm{w} = \eta \omega_\mathrm{B} R_\mathrm{B}^2$, 
with $\eta\approx 1/4$ \citep{1975A&A....39..185I}, where 
$\omega_\mathrm{B}=2\piup/P_\mathrm{B}$ is the orbital angular frequency. 
The numerical coefficient $\eta$ here can vary in a wide range due to inhomogeneities in the stellar wind and can be even negative \citep{1989MNRAS.238.1447H}; in our simulations, we varied this parameter in the range $\eta=[0,0.25]$.

The condition for quasi-spherical accretion to occur is that the specific angular momentum of a gas parcel is smaller than the Keplerian angular momentum at the Alfv\'en radius: 
$j_\mathrm{w}\le j_K(R_\mathrm{A})=\sqrt{GMR_\mathrm{A}}$. In the opposite case, $j_\mathrm{w}>j_K(R_\mathrm{A})$, the formation of an accretion disc around the magnetosphere is possible. 

The NS spin evolution at the settling accretion stage can be written in the form of angular momentum conservation equation  
\citep{2012MNRAS.420..216S,2018ASSL..454..331S}
\beq{e:spinev}
\frac{dI\omega^*}{dt}=\eta Z\dot M_\mathrm{x} \omega_\mathrm{B} R_\mathrm{B}^2-Z(1-z/Z)\dot M_\mathrm{x} \omega^*R_\mathrm{A}^2,
\eeq
where $I$ is the NS momentum of inertia, $Z=1/f(u)(u_c/u_\mathrm{ff})$ is the coupling coefficient of the plasma-magnetosphere interaction, 
$u_c/u_\mathrm{ff}\lesssim 1$ is the ratio of the convective velocity of gas in the shell and the free-fall velocity,
$z<1$  is a  numerical  coefficient  which  takes  into  account  the specific 
angular  momentum  of  the  matter. 
The variable specific angular momentum of the gravitationally captured stellar wind matter leads to the correction of the equilibrium NS period derived in 
\citet{2012MNRAS.420..216S,2018ASSL..454..331S} by the factor $0.25/\eta$. In the possible case of a negative value of $\eta$,
the NS could rapidly spin down and even start temporarily rotate in the retrograde direction. However, it is difficult to imagine that such a situation could hold much longer than the orbital binary period. Therefore,  possible episodes with negative $\eta$ would somewhat increase the NS equilibrium period 
$P_\mathrm{eq}$ which we recalculate at each time step  of  our population synthesis calculation. Below we will consider only positive values of  $\eta$ and restrict the NS spin rotation by the orbital binary period, 
$P^*\le P_\mathrm{orb}$.

It should be stressed that the hot magnetospheric shell can 
become dynamically unstable. For example, if large loops of magnetic field are present in the stellar wind, the magnetosphere can become unstable due to magnetic reconnection. This can give rise to short strong outbursts occurring in the dynamical (free-fall) time scale, such as, for example, giant flares observed in supergiant fast X-ray transients (SFXTs) 
\citep{2014MNRAS.442.2325S,2018MNRAS.474L..27H}. Stellar wind inhomogeneities (especially, variations in the wind velocity $v_\mathrm{w}$) also can disturb the settling accretion regime and even lead to free-fall Bondi accretion episodes. 
Indeed, for long-period binaries with low orbital velocities of the NS, relative variations in the mass accretion rate are related to the wind velocity and density variations by the continuity equation: 
$$
\delta \dot M_\mathrm{B}/\dot M_\mathrm{B}=\delta \rho_\mathrm{w}/\rho_\mathrm{w}-3\delta v_\mathrm{w}/v_\mathrm{w}(v_\mathrm{w}/v_{rel})^2.
$$
Wind velocity variations are most pronounced for low velocity winds with $\delta v_\mathrm{w}/v_\mathrm{w}\sim 1$, as in the case of SyXBs. Variations in the mass capture rate can give rise to significant relative variations in the X-ray luminosity, especially in the Compton cooling regime: 
$\delta L_\mathrm{x}/L_\mathrm{x}=(11/7)\delta \dot M_\mathrm{B}/\dot M_\mathrm{B}$ (see \Eq{e:MxC}).
Therefore, flaring episodes of Bondi accretion (``high'' states) on top of low-luminosity stable settling accretion intervals can be expected. This possibility should be borne in mind when comparing observational data with model calculations (see Section \ref{s:results} below).

To conclude this Section, we summarise the main changes/additions to the description of the settling accretion regime compared to our previous studies \citep{2012MNRAS.424.2265L,2015AstL...41..114K}.
\begin{enumerate}
\item Criterion of slow NS rotation for plasma entry into the NS magnetosphere via RT instability \citep{1980ApJ...235.1016A} (\Eq{e:fastP}) is applied. For faster NS rotation, accretion onto the NS occurs in the Bondi (supersonic) regime.
\item Conditions for settling accretion to occur are supplemented 
by  criteria for slow stellar winds.
\item The equilibrium NS period at the settling accretion stage takes into account variations in the specific angular momentum of the captured stellar wind (the parameter $\eta$). 
\item
Reduction of the NS equilibrium spin period due to orbital eccentricity (see Appendix A for more detail) is included.
\end{enumerate}

\section{The model}
\label{sec:method}

We use a modified version of the openly available BSE code \citep{2000MNRAS.315..543H,2002MNRAS.329..897H} appended  by the block for calculation of  spin evolution of magnetized NSs \citep{2009ARep...53..915L} 
with an account for settling accretion regime \citep[see also][for more detail]{2012MNRAS.424.2265L}.

\subsection{Formation of SyXBs}

The progenitor binary system of a SyXB with the NS accreting from stellar
wind of an evolved low-mass optical companion should harbour a primary star more
massive than $\sim 8 M_\odot$. The primary evolves off the main-sequence,
expands and can fill its Roche lobe. For a sufficiently high mass ratio, a
common envelope (CE) forms during the first mass-transfer episode (see below).
After the CE stage, the helium core of the primary evolves to a supernova that
leaves behind a NS remnant  \citep{1971ApJ...163..209W}.\footnote{ Here
and below we refer to the first numerical studies in which the formation of a bound
remnant -- a NS -- was demonstrated.} If the binary survives the supernova
explosion, a young NS appears in a binary system accompanied by a low-mass
main-sequence star. The latter, in turn, evolves off the main-sequence, and
accretion onto the NS from the stellar wind of the red giant companion becomes
possible, leading to the formation of a SyXB. 

In a narrow range of primary masses close to the lower mass limit for NS
production, after the first mass exchange the evolution of the helium core
presumably ends up with the NS formation via an electron-capture supernova
(ECSN) (\citet{1980PASJ...32..303M}; see \cite{2018A&A...614A..99S} for
the recent study), which is not accompanied by a large NS kick. We assumed that
NSs formed via ECSN get a kick of 30 km s$^{-1}$. We found, however, that
neither the mass range for ECSN nor the low kick amplitude noticeably affects
the properties of the model SyXBs. For NSs formed from Fe-core collapse, the kick
velocities of nascent NS were assumed to have a Maxwellian distribution with the
1D rms $\sigma=265$ km~s$^{-1}$ \citep{2005MNRAS.360..974H}. In any case, after
the SN explosion, the young NS finds itself in an eccentric orbit that is
gradually circularised due to the tidal interaction. The orbital circularisation
is treated as implemented in the BSE code.

In addition, we consider the possibility of formation of NSs as a result of
accretion-induced collapses (AICs) of oxygen-neon (ONe) white dwarfs which
accumulated mass close to the Chandrasekhar limit (1.4\msun) via accretion of
stellar wind \citep{1976A&A....46..229C}. Since these events are expected (i)~to
be at least by an order of magnitude less energetic than typical iron
core-collapse supernovae, (ii)~to have a  very low amount of mass lost, 
and (iii)~to have a small collapse anisotropy
\citep{2006ApJ...644.1063D}, we assign to the NSs produced
via this channel kick velocities $30\,\kms$, like for descendants of ECSN.

\subsection{Common envelope stage}
 
Common envelope is the most prominent example of non-conservative binary
evolution enabling an efficient angular momentum removal from the binary on a
short time-scale (see e.g. \cite{2014LRR....17....3P,2017MNRAS.464.1607Y} for
more detail and references). Many details of the binary evolution during the CE
stage remain unclear, but it is widely accepted to parameterize the CE evolution
by the so-called $\alpha$-formalism \citep{web84}, in which the dimensionless
parameter $\alpha_\mathrm{CE}$ means the fraction of the orbital energy
difference between the onset and the end of the CE stage that is spent to unbind
the envelope of the evolved (post main-sequence) companion $M_1$ from its core
$M_c$:
\beq{}
\alpha_\mathrm{CE}\left(-\frac{GM_1M_2}{2a_i}+\frac{GM_cM_2}{2a_f}\right)=-\frac{G(M_1-M_c)M_1}{\lambda R_1}.
\label{eq:ce}
\eeq
Here the second star $M_2$ is assumed to conserve mass during the CE stage, and the binding energy of 
the stellar envelope is parameterized by the dimensionless parameter $\lambda$ (see calculations for 
stars of different metallicities in \cite{2011ApJ...743...49L}). This model suggests that the product 
$\alpha_\mathrm{CE}\lambda$ determines the ratio of the final to initial binary separations $a_f/a_i$ 
after the CE stage: the smaller 
$\alpha_\mathrm{CE}\lambda$ (i.e., ``more efficient CE''), the smaller 
$a_f/a_i$. 

Alternatively, it was also suggested that the first 
CE-stage of
evolution can be controlled by the system angular momentum $J$ conservation (the so-called
$\gamma$-formalism that implicitly assumes energy conservation
\citep{2000A&A...360.1011N}):
\begin{equation}
\frac{\Delta J}{J} = \gamma \frac{\Delta M}{M}\,,
\label{eq:gammaform}
\end{equation}
where $M$ is total mass of the system. Then, the CE evolution can be not accompanied
by strong reduction of separation between the binary components. \citet{2005MNRAS.356..753N}
have shown that the $\gamma$-formalism is able to explain the formation of a variety of
binary systems, including low-mass X-ray binaries, if $\gamma$ is confined to a
rather narrow range 1.5 -- 1.75. If one assumes $\gamma$=1.75, the post-CE
separation between the binary components is close to that obtained for $\alpha_\mathrm{CE}$=4
(for fixed $\lambda=0.5$), i.e. formally requires additional energy sources,
apart from the orbital energy, for expulsion of CE. Clearly, the parameter
$\alpha_\mathrm{CE}$ can (and probably should) vary from system to system and
with evolutionary stage of the donor, but in the absence of a firm theory and
any realistic numerical simulations, in our calculations we will treat
$\alpha_\mathrm{CE}$ as a free parameter, one and the same for all binaries, that can
take values both less and higher than unity (see also the discussion in
\citet{2014LRR....17....3P,2017MNRAS.464.1607Y}).

\subsection{Stellar winds from 
evolved low-mass stars}
\label{s:wind}

The wind mass-loss from red-giants is calculated using the formula from \cite{1978A&A....70..227K} with the parameter $\eta_\mathrm{w}=0.5$. Stellar wind velocities from red giants and AGB-stars are described as in 
\citet[][(Eqs.~(13)-(17), the terminal wind velocity  5~km~s$^{-1}$]{2006MNRAS.372.1389L}). Stellar winds from stripped helium 
stars are treated according to \citet{2017A&A...607L...8V}. 

SyXBs usually have moderately non-circular orbits. For low wind velocities, the most important accretion effects are due to the NS orbital velocity variations. In the population synthesis simulations, the orbital eccentricity effects in stellar winds  can be taken into account, for example, by averaging the relevant quantities over the orbital period of the binary. For noticeable orbital eccentricities this results in a substantial decrease of the NS equilibrium periods at the settling accretion stage  \citep{2018arXiv181102842P}.  For low eccentricities 
($e\lesssim 0.2$) the eccentricity effects are found to be not very significant (see Appendix A for more detail). 

\subsection{Conditions for settling accretion}

The typical NS spin evolution proceeds as follows (see \citet{2009ARep...53..915L} for more details and \citet{2012MNRAS.424.2265L} for the description of  NS spin evolution in SyXBs). Initial NS spin periods are assumed to follow a Gaussian distribution centered at $P_0=100$~ms with comparable dispersion 
\citep{2016MNRAS.463.1642P}. Magnetic fields of NS are assumed to be distributed log-normally, $f(\mu)d\mu\sim \exp[-(\log\mu-\log\mu_0)^2/\sigma_\mu^2]$, with the mean value $\log\mu_0=30.35$ and dispersion $\sigma_\mu=0.55$. Accretion-induced magnetic field decay is taken into account as in \citet{2012MNRAS.424.2265L}. Initially,  the NS is in the ``ejector'' stage when the pressure of the relativistic wind generated near the NS surface dominates that of the surrounding plasma. Then the 
NS spins down to the ``propeller'' stage in which the centrifugal forces at the magnetospheric boundary prevent matter accretion. The propeller stage terminates 
when the corotation radius $R_c=(GM_\mathrm{x}P^{*2}/4\piup^2)^{1/3}$ becomes larger than the magnetosphere radius $R_\mathrm{A}$, $R_c \ge R_\mathrm{A}$, enabling accretion.

In the binary systems without Roche-lobe overflow, we check which type of accretion onto the NS is possible -- disc or quasi-spherical one. 
If the condition for the spherical accretion 
($j_\mathrm{w}<j_K(R_\mathrm{A})=\sqrt{GM_\mathrm{x}R_\mathrm{A}}$) is met, we 
check the conditions for the settling accretion to occur:

(i) Sufficiently slow rotation of the NS enabling RT instability to provide the plasma entry into the NS magnetosphere (see \Eq{e:fastP} above). For faster spinning NS, plasma entry rate into the magnetosphere is regulated by the Kelvin-Helmholtz instability, and free-fall Bondi accretion is realised \citep{1983ApJ...266..175B}.

(ii) Average X-ray luminosity is below $L^\dag\simeq 4\times 10^{36}$~erg s$^{-1}$ to prevent rapid plasma cooling.

(iii) If the relative wind velocity $v_\mathrm{rel}$ is below $\simeq 80$ km s$^{-1}$, the photoionization heating of gas above the Bondi radius $R_\mathrm{B}$ up to $T_{max}\simeq 5\times 10^5$~K is significant \citep{2012MNRAS.420..216S}, no strong bow shock is formed near $R_\mathrm{B}$, and the accretion proceeds from the effective Bondi radius $R_\mathrm{B}^*\simeq 3.5\times 10^{12}$~cm. Therefore, for low wind velocities and wide binary systems, when the Bondi radius calculated by formula \eqn{e:RB} can be formally very large, we set 
$R_\mathrm{B}=\min(R_\mathrm{B}, 50R_\odot)$.

In binaries with Roche-lobe overflow, only disc accretion takes place.

\subsection{Initial distributions}

The  masses of 
primary stars are distributed according to the Salpeter law, 
$dN/dM_1\propto M_1^{-2.35}$ ($0.1 M_\odot\le M_1\le 100 M_\odot$). 
Orbital separations of binaries  are distributed following 
\cite{2012Sci...337..444S}.  
Binary mass ratios $q=M_2/M_1\le 1$ and orbital eccentricities 
$e$ are assumed to have flat distributions: $dN/dq$=const., 
$dN/de$=const. in the range [0,1]. We assume that Galactic stellar
binarity rate is 50\%, i.e. 2/3 of stars enter binary systems.

\subsection{Galactic star-formation history}

To compare the population synthesis results with observations, we convolved the formation rate of SyXB with the Galactic star formation rate (SFR).  
We adopted a star-formation history in the Galactic bulge and thin disc in the form suggested by \cite{2010A&A...521A..85Y}:
\beq{e:SFR}
\frac{\mathrm{SFR}(t)}{M_\odot\,\mathrm{yr}^{-1}}=\left\{
\begin{array}{lr}
11e^{-\frac{t-t_0}{\tau}}+0.12(t-t_0), & t\ge t_0\\
0, & t<t_0
\end{array}
\right.
\eeq
with time $t$ in Gyr, $t_0$=4 Gyr, $\tau=9$~Gyr, Galactic age 14 Gyr. This model gives a total mass of the Galactic bulge and thin disk 
$M_G=7.2\times 10^{10} M_\odot$, which we use for the normalisation of our calculations. 
\begin{figure}
         \includegraphics[width=0.49\textwidth]{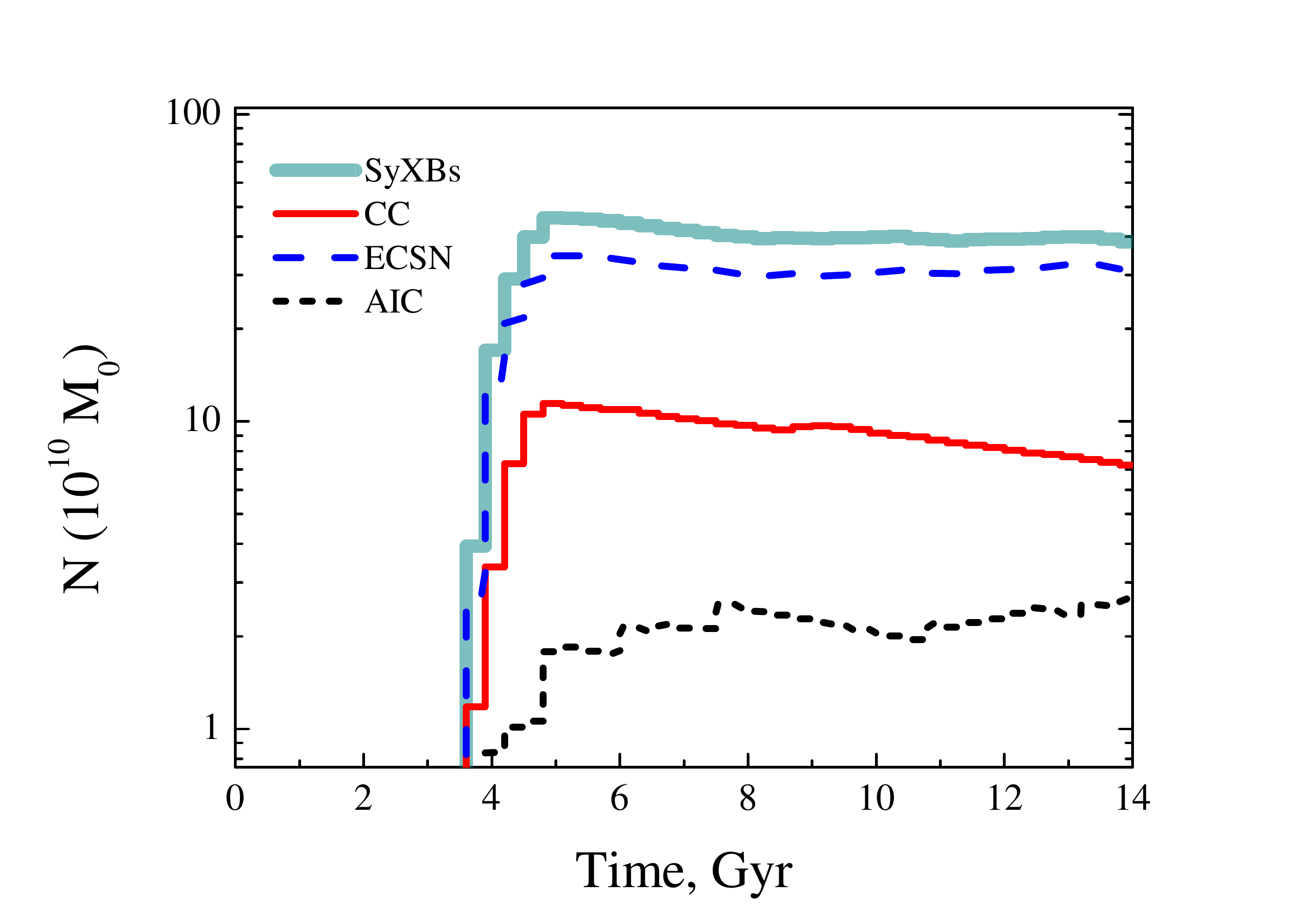}
\caption{Galactic number of SyXBs as a function of time for SyXBs with the NS-components produced 
via iron core-collapse (CC), 
ECSN, and AIC. Thick solid line at the top shows total number of SyXBs.
The systems with FGB-donors are omitted since they are not symbiotic stars in a
strict sense. For descendants of AICs, we present the upper estimate of their
number obtained by assuming 100\% efficient accretion.
}
\label{f:rate}
\end{figure}
\begin{figure*}    
	\includegraphics[width=0.495\textwidth]{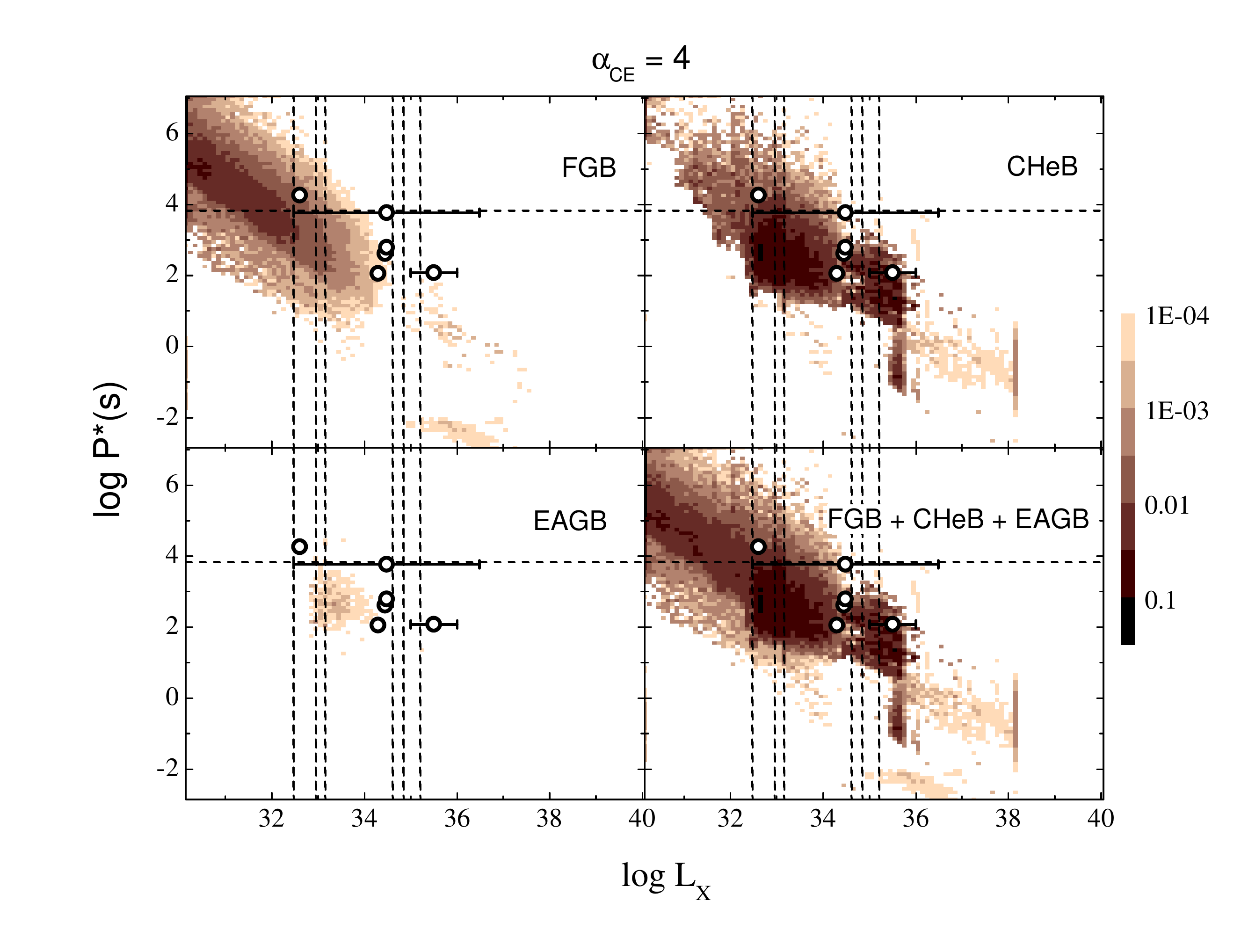}
    \includegraphics[width=0.495\textwidth]{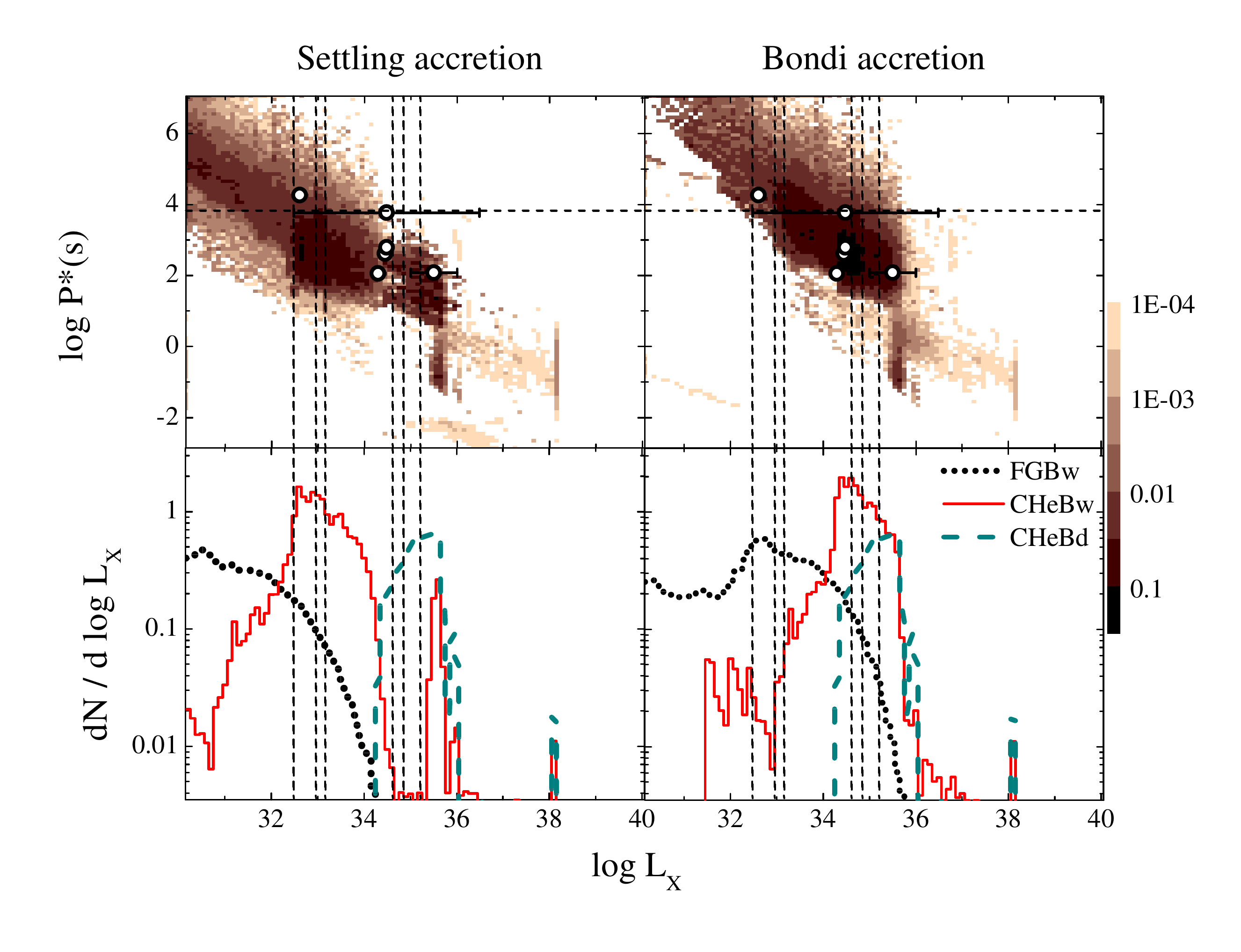}
    \caption{Left: NS spin period $P^*$ -- X-ray luminosity 
$L_\mathrm{x}$ diagram  for accreting NS with optical companions at different evolutionary 
stages: FGB -- first giant branch, CHeB -- core helium burning, EAGB -- early AGB stage, calculated for 
the common envelope parameter 
$\alpha_\mathrm{CE}=4$, the envelope binding energy parameter 
$\lambda$
after \citet{2011ApJ...743...49L}  and flat initial mass ratio distribution $dN/dq$=const.  
Known SyXBs from Table 1 are shown by open circles with error bars or by dashed lines, if only $L_\mathrm{x}$ or $P^*$ is known. In grey scale 
shown is the number density of 
model SyXBs per $10^{10}$\msun\ (see text). 
Right: $P^*-L_\mathrm{x}$ diagram for wind-accreting (FGBw, CHeBw) and disc-accreting  (CHeBd) systems (upper row) and their differential luminosity function 
$dN/d\log L_\mathrm{x}$ (bottom row). Left and right columns correspond to quiet settling accretion luminosity $L_\mathrm{x}$ and maximum possible outburst (Bondi) X-ray luminosity $L_{\mathrm{x,B}}$, respectively.
 We remind that, strictly speaking, systems with FGB-donors can not be considered as SyXBs and are shown in this Figure and Figure~\ref{f:porblx}  only for completeness, as an example of model population of wind-fed LMXB subject to settling accretion, like SyXBs. Note that known SyXBs are not found in the regions of the diagrams occupied by this population. 
}
 \label{f:plx}	
\end{figure*}

\begin{figure*}
	\includegraphics[width=0.49\textwidth]{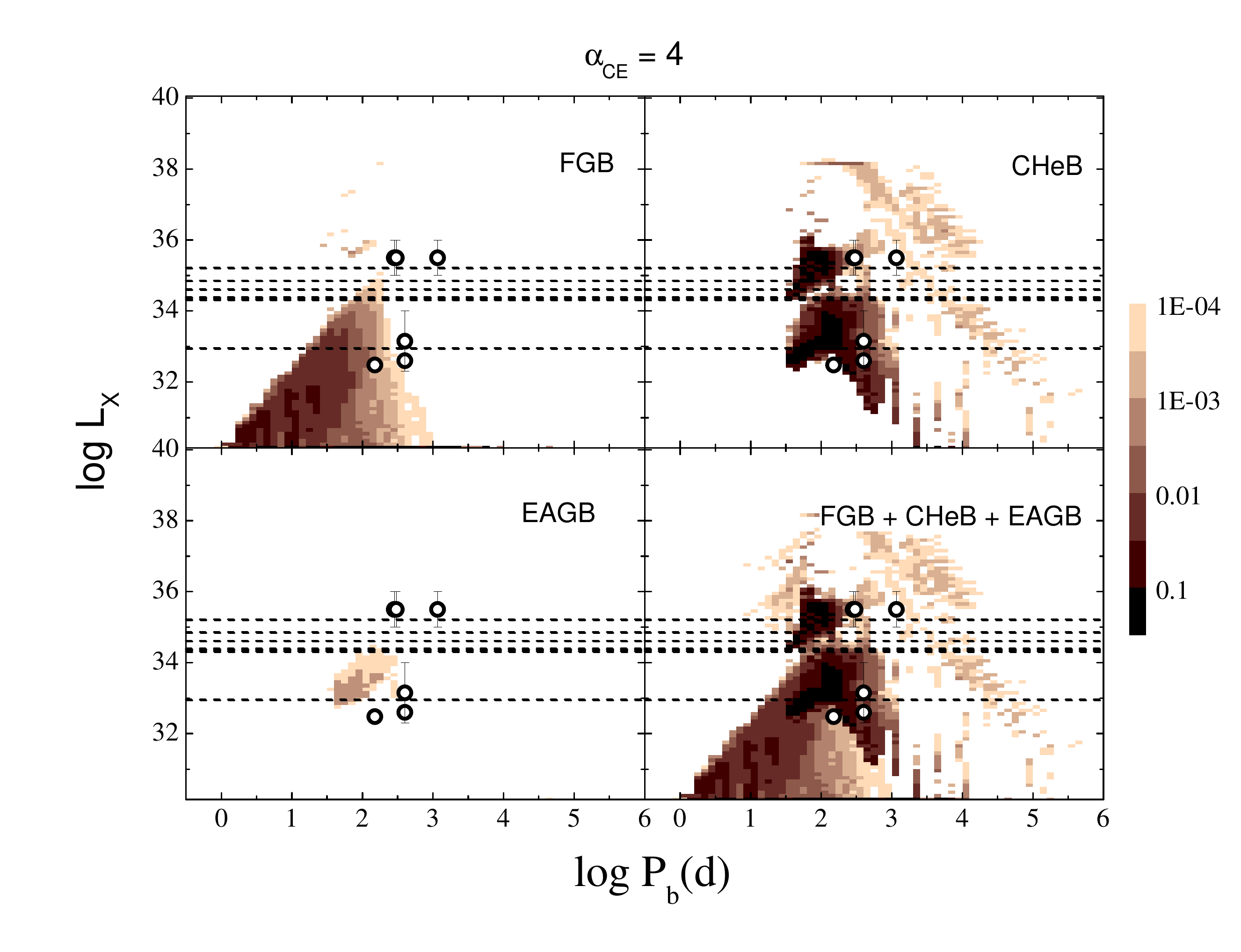}
    \includegraphics[width=0.49\textwidth]{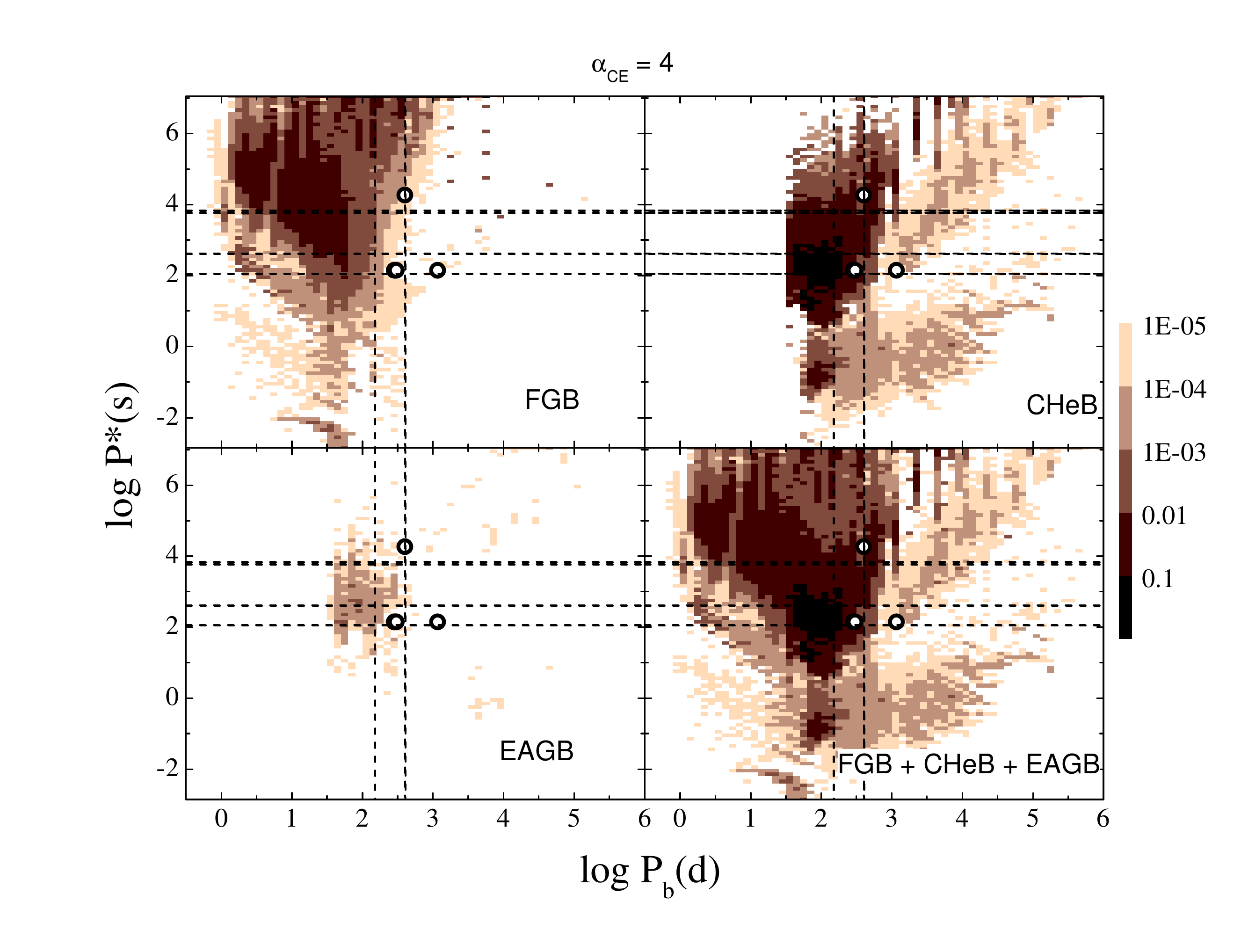}
    \caption{Model $L_\mathrm{x}-P_\mathrm{b}$ diagram (left panel) and
the Corbet diagram $P^*-P_\mathrm{B}$ for  wind- and disc-accreting systems from Figure \ref{f:plx}. Open symbols show the known SyXBs.  Model parameters as in Fig. \ref{f:plx}. 
 }   
 \label{f:porblx}
\end{figure*}


\section{Results}
\label{s:results}

In our simulations, we have run $5\times 10^6$ binary systems for different
values of the parameter $\alpha_\mathrm{CE}$ that mostly influences the number
of SyXBs and their observed characteristics --- NS spin period at the
wind-accretion stage $P^*$, X-ray luminosity $L_\mathrm{x}$ and binary orbital
period $P_\mathrm{B}$. The results for $\alpha_{CE} = 4$, which provides the
best agreement with observations, are presented in 
Figs.~\ref{f:rate} -- \ref{f:porblx}. 

\subsection{Galactic number of SyXBs}

 In our calculations, we selected \textit{all} detached systems with the NS
components that have companions that reached the first red giant branch (FGB) or
are in the later stages of the evolution (core He-burning stage -- CHeB,
early-AGB stars -- EAGB), but do not overflow their Roche lobes. By virtue of
such a selection, our sample contains a number of binaries with orbital periods well
below several hundred day (see Fig.~\ref{f:porblx}). Strictly speaking, such
objects do not comply with the ``standard'' definition of symbiotic stars that 
defines them as systems with donors -- late giants \citep[e.g.,
][]{1984ApJ...279..252K}, because in binaries with such short orbital periods there is no room for late giants. Examination of
the distribution over orbital periods of about 100 symbiotic stars with white
dwarf components with known \porb\ \citep[][Fig.~1]{2012BaltA..21....5M} tells that it is
justified to consider as SyXB systems only those with \porb$\apgt$200\,day. This
suggests us to consider NS+FGB systems as a separate subclass of wind-accreting
low-mass X-ray binaries\footnote{Note that these objects have, predominantly,  hardly detectable 
low $L_\mathrm{x}$.} and leaves as SyXBs predominantly
systems with CHeB components. For this reason, we omit NS+FGB stars in
Fig.~\ref{f:rate} that shows the evolution of the number of Galactic SyXBs with time, 
calculated by convolution of the SyXB formation delay-time distribution ${\cal
R}(t)$ with the assumed SFR rate (\Eq{e:SFR}):
$N(t)=\int_0^t\mathrm{SFR}(t-\tau){\cal R}(\tau)d\tau$. Note that even not all
CHeB-donor systems fit definition of SyXBs. Symbiotic X-ray binaries with CHeB
components form, typically, in less than 1.5\,Gyr after the formation. Therefore, their Galactic 
number evolution simply follows SFR set by Eq.~(\ref{e:SFR}).    
 
Figure~\ref{f:rate} shows that most of the NSs in SyXBs originate from ECSNs. The reason for this is clear: in their precursors, companions of
the NS progenitors are (1-2)\,\ms\ stars. In such a case, a natal kick of only
$\sim100\kms$ is sufficient to disrupt the binary in the SN event.   

In Fig.~\ref{f:rate}, we also present the number of SyXBs with NS-components
descending from white dwarfs via accretion-induced collapses, which is obtained assuming a 100\%
efficiency of accretion onto WD. Evidently, this number is an upper limit and can very
strongly overestimate the real number of such systems. One of the main reasons is low accretion efficiency due to
the well known mismatch between the ranges of accretion rates allowing stable
hydrogen and helium burning, which prevents effective growth of mass of the accreting WD
\citep[see, e.g., ][]{1996ApJS..105..145I}. As yet, it is impossible to quantify
the effect of retention of matter accreted by ONe WDs (which, in principle, may
be even negative) in the population synthesis code because of the absence of
systematic computations for grids of models in the ($M_{\rm WD}, \dot M_{\rm
accr}$) space. Extrapolation of the extant data for the models of CO WDs may result
in systematic errors, since for ONe WDs the masses of the critical shells of
hydrogen needed for explosive burning are about twice as large as for CO WDs
\citep{1998ApJ...494..680J,2003A&A...407.1021G}, hence the outbursts may be less
frequent but more violent and to eject more mass. As follows from the estimates of
\citet{1996ApJS..105..145I} and calculations of \citet{2014ApJ...794...28P},
efficient accumulation of matter by massive WDs in symbiotic systems is possible
for accretion rates $\sim 10^{-7} - 10^{-8}$\,\myr. However, such a relatively
high $\dot M_{\rm accr}$ is not feasible for about a half of symbiotic stars, if
accretion occurs from stellar wind \citep{2006MNRAS.372.1389L}. Even if the
mass of an ONe WD attains \mch, it is not clear as yet, whether the
collapse will result in the formation of a NS or in the white dwarf destruction. According to
\citet{2015MNRAS.453.1910S,2016A&A...593A..72J}, the collapse occurs if oxygen
deflagration starts at the central density not lower than about $10^{10} 
\mathrm {g\,cm^{-3}}$. The collapse may be prevented by ignition of residual $^{12}$C which may
be present in WD in non-negligible amount \citep{2018arXiv181013002S}. Finally,
we note that most of the 
model systems with NSs
resulting from AICs have orbital periods too short to be identified with
symbiotic stars. This is in qualitative agreement with
\citet{2010MNRAS.402.1437H} who have shown, using treatment of
retention implemented in BSE, that wide systems harbouring post-AIC NSs are exceptionally
rare. Below we will omit post-AIC NSs from consideration because the modeling of their
magnetic fields and initial spins requires the study of evolution of their WD progenitors, which 
is beyond the scope of the present paper, and their vanishingly small number.   

To summarise, we can expect that the current number of SyXBs in the Galaxy is slightly less than 40 
objects per $10^{10}$\,\ms.

\subsection{NS spin period-luminosity distribution}

Figure \ref{f:plx}, left panel, shows the $P^*-L_\mathrm{x}$ distribution of
model wind-accreting NSs with evolved companions with masses below 
3 $M_\odot$. Here $L_\mathrm{x}$ is the X-ray luminosity at the settling
accretion stage. In the grey scale shown is the model number of systems 
normalized to $10^{10}$\,\msun. Observed SyXBs from Table 1 with known spin
periods and X-ray luminosities are shown by open circles with error bars. Positions of
the systems with known luminosities or periods only are indicated by thin dashed
lines. 
Wind-fed low-mass X-ray binaries  with optical components of different types 
are shown separately, from left to right, top to
bottom: NS with optical star at the first giant branch (FGB), NS with companion
in the core He burning stage (CHeB) and NS with companion at the early AGB stage
(EAGB).  As discussed above, mostly, the population with CHeB-components may be identified with SyXBs.
In the right bottom panel, three sub-populations are over-plotted. In
Fig.~\ref{f:plx}, several regions are clearly seen. The most populated region is
occupied by the model systems at the settling accretion stage with the mean spin
periods 100--10000~s and $L_\mathrm{x}\sim 10^{32}-10^{34}$ erg s$^{-1}$. The
population of systems with $P^*\lesssim 30$~s best seen in the rightmost bottom
panel in the plot at the left corresponds to NSs on the Bondi accretion stage.
The strip at the bottom of the plot at $P^*\sim 10$~ms is formed by disc-accreting millisecond
recycled NSs. The vertical strip at $L_\mathrm{x}\sim 10^{38}$~erg s$^{-1}$ is
populated by super-Eddington accreting NSs. The last two types of accreting NSs should not
be related to SyXBs, but they are descendants of SyXBs and are shown for
completeness.

The right panel of Fig.~\ref{f:plx} shows
the populations of  stars  with FGB and CHeB components in the regimes of settling accretion and 
disc-accretion and the corresponding differential X-ray luminosity functions $dN/d\log L_\mathrm{x}$. For the 
latter systems, $L_\mathrm{x}$ corresponds to the outbursts, i.e., represents the maximum of
$L_\mathrm{x}$.  

Note also that for the $\alpha$-formalism of common envelopes, even if we would
assume that the entire orbital energy is spent to expel the CE
($\alpha_\mathrm{CE}=1$), most of the model binaries would have parameters different
from those of observed systems. However, the position of the model systems in the
$P^*-L_\mathrm{x}$ diagram well overlaps with that of observed ones if the
$\gamma$-formalism with parameter 1.75 is applied for the CE stage (see also
Fig.~\ref{f:plx}, right panel). Formally, this is equivalent to
$\alpha_\mathrm{CE} \approx 4$ and may mean that energy sources other than the binary
orbital energy are necessary for expulsion of the CE (see also
\citet{2017MNRAS.464.1607Y} and the discussion of low CE efficiency found in
3D-simulations by \citet{2018arXiv181103656R}). 

Figure \ref{f:plx} suggests that systems with  CHeB donors mostly contribute
to the Galactic population of SyXBs; the number of SyXBs with EAGB donors
is insignificant because of the short duration of this stage. Generally, the
location of the observed SyXBs from Table 1 in this diagram is in agreement with
calculations. We also stress here that the intrinsically unstable character of
the wind subsonic accretion we discussed above enables the model systems to
shift by an order of magnitude in the X-ray luminosity up to the maximum
available value for the Bondi-Hoyle wind accretion $L_{\mathrm x,B}$, which would result
in better agreement with observations (see the upper rows of right panels in Fig. \ref{f:plx}). 

\subsection{X-ray luminosity -- orbital period distribution}

The left panel of Fig.~\ref{f:porblx} shows the model
$L_\mathrm{x}-P_\mathrm{b}$ distribution of wind-accreting systems. As in the case of
$P^*-L_\mathrm{x}$ plot, the observed systems (shown by open circles and
dashed lines) are better reproduced by the model with low-efficiency CE (large
$\alpha_\mathrm{CE}=4$). A more efficient CE would give
rise to model SyXBs with shorter orbital periods than observed.  

There is a pronounced feature in the $L_\mathrm{x}-P_\mathrm{B}$ plot, a region
at short orbital periods and low X-ray luminosities, quite densely populated by
model systems, which persists for all model parameters. These model sources are
produced by NS+FGB systems (see leftmost upper panel of Fig.~\ref{f:porblx}). 
There can be also a small contribution of high-luminosity long-period NS+CHeB
SyXBs (see Fig.~\ref{f:plx}). 
However, as discussed above, we do not identify these systems with SyXBs. Presently, no wind-fed LMXBs  
with such parameters are
known. Detection of such sources by future more sensitive X-ray observations
would provide an important test of the model.

\subsection{NS spin period -- orbital period distribution (the Corbet diagram)}

The model Corbet diagrams ($P^*-P_\mathrm{b}$) are presented in the right panel
of Fig.~\ref{f:porblx}, for the same model parameters as in Fig. \ref{f:plx}.
Like Fig. \ref{f:plx}, the model predicts appearance of a sub-population of
rapidly rotating disc-accreting NSs (recycled millisecond X-ray pulsars), which are descendants of SyXBs. In addition,
very slowly rotating low-luminosity NSs are also present in this diagram. Clearly, the 
observed SyXBs from Table 1 fall within model regions on this plot.
However, small number statistics prevents us from obtaining firm conclusions. 

\section{Discussion and conclusions}
\label{s:disc}
In this paper, we present the results of population synthesis of Galactic
symbiotic X-ray binaries harbouring a neutron star accreting from the wind of
low-mass evolved companion. We have used a modified version of the open
population synthesis code BSE supplemented by a detailed treatment of the
evolution of rotating magnetized neutron stars. In comparison to our previous
studies \citep{2012MNRAS.424.2265L,2015AstL...41..114K}, the code was upgraded
for an accurate treatment of different regimes of wind accretion (subsonic and
supersonic quasi-spherical and disc regimes). This enabled us to reach better
consistency of the model with observed characteristics of known SyXBs. 

We have found that under the usual assumptions about the binary evolution and NS formation parameters, the model NS spin period -- X-ray luminosity (Fig. \ref{f:plx}, left panel), X-ray luminosity -- orbital period (Fig. \ref{f:porblx}, left panel) and NS spin period -- orbital period (Fig. \ref{f:porblx}, right panel) distributions are {\em simultaneously} in agreement with observations if we assume a low efficiency of matter expulsion in the common envelope stage characterized by the CE efficiency parameter $\alpha_\mathrm{CE}=4$, suggesting that other energy sources  than only orbital energy of the companions 
are required to expel the common envelope. This finding is in line with independent indications of low CE efficiency reported in \citet{2016ApJ...816L...9O,2018arXiv181103656R} and discussed in 
\citet{2017MNRAS.464.1607Y} for  possible progenitors of SNe~Ia. Other evolutionary parameters (mass range of progenitors of NSs formed via electron-capture supernovae and the value of the NS kick velocity, initial NS magnetic field distribution and parameters of the NS field decay, etc.) are found to have minor effect on the model properties of SyXBs. 

We also stress that the observed X-ray luminosity of SyXBs is better reproduced when the model systems are at high (flaring) stage, with the mass accretion rate onto NS being equal to its maximum possible (Bondi) value $\dot M_\mathrm{B}$ from the stellar wind of the companion. This fact has a clear physical explanation: while the settling accretion stage (with reduced mass accretion rate $\dot M_\mathrm{x}=f(u)\dot M_\mathrm{B}$, see \Eq{e:Mxrad}) seems to be unavoidable to explain the long spin periods of accreting NSs and observed correlations, e.g. in GX 1+4 \citep{GonzalezGalan2011,2012MNRAS.420..216S}, settling accretion becomes progressively unstable in the radiative cooling regime. That is, at $\dot M_\mathrm{B}\ll 4\times 10^{36}$ erg s$^{-1}$, the source should demonstrate a flaring behaviour, with the mass accretion rate reaching 
$\dot M_\mathrm{B}$ in outbursts. An extreme example of such a behaviour could be supergiant soft X-ray transients (SFXTs) \citep{2014MNRAS.442.2325S}, with short duty-cycle of outbursts. However, in the case of SyXBs, the properties of stellar winds are very different from those of early type supergiants in SFXTs, and the outburst duty cycle can be longer (e.g., as in GX 1+4). Therefore, it is more likely to observe a SyXB in the flaring state.

The Galactic number of model SyXBs was calculated using the convolution of the population calculated for an instantaneous star formation burst with  the Galactic star-formation rate history  (see Fig. \ref{f:rate}). For the adopted Galactic SFR, this number is 
 slightly below 40  per $10^{10}$ \msun\ and was almost constant during the last 10~Gyr. Of all initial binary parameters, it is
mostly affected by the shape of the initial binary mass ratio distribution and decreases by factor of a few when changing from a flat mass ratio distribution  to that peaked at equal initial masses of binary components.  In our model, about 80\% of the NS components of SyXBs are formed by electron-capture supernovae and the rest -- by the iron-core collapse SN. The number of
SyXBs with the post-accretion-induced-collapse NSs is found to be vanishingly small. We expect that the majority of wind-emitting components in SyXBs should be core-helium burning stars.

The computed number of currently existing Galactic SyXBs (Fig.~\ref{f:rate}) is
found to be  substantially lower than predicted in our previous studies and is more consistent with the low number of observed objects, though, still exceeding the latter. We should note, however, after  
\citet{2012A&A...538A.123M}, that identification of new SyXBs requires a very precise (smaller than a few arc sec) localisation of X-ray sources in order to find their optical counterparts, which is a hard task in the crowded star fields, especially in the Galactic bulge. 

The model also predicts  the existence of  very slowly rotating NSs in SyXBs, which spin down at the low-luminosity settling accretion stage. Detailed studies of the X-ray properties of Galactic SyXBs can be used as an independent instrument to test the evolution of binary stars in the Galaxy. We expect that the forthcoming all-sky X-ray survey like the SRG-eROSITA mission 
\citep{2016SPIE.9905E..1KP,2018SPIE10699E..5HP} will be able to discover a dozen of Galactic SyXBs similar to the objects  described in the present paper. 

 Our calculations predict the existence of a population of wind-fed LMXB also subject to settling accretion,
which hosts FGB stars in the systems with orbital periods below about 200 -- 300 days. This range of periods is shorter than the one typical for ``classical'' symbiotic binaries. Most of them are also X-ray dim sources with 
$L_\mathrm{x} \aplt 10^{32}\,\mathrm{erg~s^{-1}}$, lower than $L_\mathrm{x}$ of typical known SyXBs. We expect that these 
binaries in the course of evolution will form common envelopes upon Roche lobe overflow by the optical  components and merge or turn into  NS+WD systems.

\section*{Acknowledgements}

We thank the anonymous referee for useful remarks. This study was partially supported by RFBR grant No.~19-02-00790.
 The work of AGK is supported by RSF grant No.~14-12-00146 (population synthesis calculations).
The work of KAP (Section 2 and Appendix A) is partially  supported by RFBR grant 18-502-12025. KAP and AGK also acknowledge the support from the Program of development of M.V. Lomonosov Moscow State University (Leading Scientific School 'Physics of stars, relativistic objects and galaxies').


\bibliographystyle{mnras}
\bibliography{./paper}

\begin{thebibliography}{}
\makeatletter
\relax
\def\mn@urlcharsother{\let\do\@makeother \do\$\do\&\do\#\do\^\do\_\do\%\do\~}
\def\mn@doi{\begingroup\mn@urlcharsother \@ifnextchar [ {\mn@doi@}
  {\mn@doi@[]}}
\def\mn@doi@[#1]#2{\def\@tempa{#1}\ifx\@tempa\@empty \href
  {http://dx.doi.org/#2} {doi:#2}\else \href {http://dx.doi.org/#2} {#1}\fi
  \endgroup}
\def\mn@eprint#1#2{\mn@eprint@#1:#2::\@nil}
\def\mn@eprint@arXiv#1{\href {http://arxiv.org/abs/#1} {{\tt arXiv:#1}}}
\def\mn@eprint@dblp#1{\href {http://dblp.uni-trier.de/rec/bibtex/#1.xml}
  {dblp:#1}}
\def\mn@eprint@#1:#2:#3:#4\@nil{\def\@tempa {#1}\def\@tempb {#2}\def\@tempc
  {#3}\ifx \@tempc \@empty \let \@tempc \@tempb \let \@tempb \@tempa \fi \ifx
  \@tempb \@empty \def\@tempb {arXiv}\fi \@ifundefined
  {mn@eprint@\@tempb}{\@tempb:\@tempc}{\expandafter \expandafter \csname
  mn@eprint@\@tempb\endcsname \expandafter{\@tempc}}}

\bibitem[\protect\citeauthoryear{{Arons} \& {Lea}}{{Arons} \&
  {Lea}}{1976}]{1976ApJ...207..914A}
{Arons} J.,  {Lea} S.~M.,  1976, \mn@doi [\apj] {10.1086/154562}, \href
  {http://adsabs.harvard.edu/abs/1976ApJ...207..914A} {207, 914}

\bibitem[\protect\citeauthoryear{{Arons} \& {Lea}}{{Arons} \&
  {Lea}}{1980}]{1980ApJ...235.1016A}
{Arons} J.,  {Lea} S.~M.,  1980, \mn@doi [\apj] {10.1086/157706}, \href
  {http://adsabs.harvard.edu/abs/1980ApJ...235.1016A} {235, 1016}

\bibitem[\protect\citeauthoryear{{Bahramian}, {Gladstone}, {Heinke},
  {Wijnands}, {Kaur}  \& {Altamirano}}{{Bahramian}
  et~al.}{2014}]{2014MNRAS.441..640B}
{Bahramian} A.,  {Gladstone} J.~C.,  {Heinke} C.~O.,  {Wijnands} R.,  {Kaur}
  R.,   {Altamirano} D.,  2014, \mn@doi [\mnras] {10.1093/mnras/stu611}, \href
  {http://adsabs.harvard.edu/abs/2014MNRAS.441..640B} {441, 640}

\bibitem[\protect\citeauthoryear{{Bozzo} et~al.,}{{Bozzo}
  et~al.}{2018}]{2018A&A...613A..22B}
{Bozzo} E.,  et~al., 2018, \mn@doi [\aap] {10.1051/0004-6361/201832588}, \href
  {http://adsabs.harvard.edu/abs/2018A%26A...613A..22B} {613, A22}

\bibitem[\protect\citeauthoryear{{Burnard}, {Arons}  \& {Lea}}{{Burnard}
  et~al.}{1983}]{1983ApJ...266..175B}
{Burnard} D.~J.,  {Arons} J.,   {Lea} S.~M.,  1983, \mn@doi [\apj]
  {10.1086/160768}, \href {http://adsabs.harvard.edu/abs/1983ApJ...266..175B}
  {266, 175}

\bibitem[\protect\citeauthoryear{{Canal} \& {Schatzman}}{{Canal} \&
  {Schatzman}}{1976}]{1976A&A....46..229C}
{Canal} R.,  {Schatzman} E.,  1976, \aap, \href
  {http://esoads.eso.org/abs/1976A%26A....46..229C} {46, 229}

\bibitem[\protect\citeauthoryear{{Chakrabarty} et~al.,}{{Chakrabarty}
  et~al.}{1997}]{1997ApJ...481L.101C}
{Chakrabarty} D.,  et~al., 1997, \mn@doi [\apjl] {10.1086/310666}, \href
  {http://adsabs.harvard.edu/abs/1997ApJ...481L.101C} {481, L101}

\bibitem[\protect\citeauthoryear{{Corbet}, {Sokoloski}, {Mukai}, {Markwardt}
  \& {Tueller}}{{Corbet} et~al.}{2008}]{2008ApJ...675.1424C}
{Corbet} R.~H.~D.,  {Sokoloski} J.~L.,  {Mukai} K.,  {Markwardt} C.~B.,
  {Tueller} J.,  2008, \mn@doi [\apj] {10.1086/526337}, \href
  {http://adsabs.harvard.edu/abs/2008ApJ...675.1424C} {675, 1424}

\bibitem[\protect\citeauthoryear{{Davidsen}, {Malina}  \& {Bowyer}}{{Davidsen}
  et~al.}{1977}]{1977ApJ...211..866D}
{Davidsen} A.,  {Malina} R.,   {Bowyer} S.,  1977, \mn@doi [\apj]
  {10.1086/154996}, \href {http://adsabs.harvard.edu/abs/1977ApJ...211..866D}
  {211, 866}

\bibitem[\protect\citeauthoryear{{Dessart}, {Burrows}, {Ott}, {Livne}, {Yoon}
  \& {Langer}}{{Dessart} et~al.}{2006}]{2006ApJ...644.1063D}
{Dessart} L.,  {Burrows} A.,  {Ott} C.~D.,  {Livne} E.,  {Yoon} S.-C.,
  {Langer} N.,  2006, \mn@doi [\apj] {10.1086/503626}, \href
  {http://esoads.eso.org/abs/2006ApJ...644.1063D} {644, 1063}

\bibitem[\protect\citeauthoryear{{Elsner} \& {Lamb}}{{Elsner} \&
  {Lamb}}{1984}]{1984ApJ...278..326E}
{Elsner} R.~F.,  {Lamb} F.~K.,  1984, \mn@doi [\apj] {10.1086/161797}, \href
  {http://adsabs.harvard.edu/abs/1984ApJ...278..326E} {278, 326}

\bibitem[\protect\citeauthoryear{{Ferrigno}, {Segreto}, {Santangelo}, {Wilms},
  {Kreykenbohm}, {Denis}  \& {Staubert}}{{Ferrigno}
  et~al.}{2007}]{2007A&A...462..995F}
{Ferrigno} C.,  {Segreto} A.,  {Santangelo} A.,  {Wilms} J.,  {Kreykenbohm} I.,
   {Denis} M.,   {Staubert} R.,  2007, \mn@doi [\aap]
  {10.1051/0004-6361:20053878}, \href
  {http://adsabs.harvard.edu/abs/2007A%26A...462..995F} {462, 995}

\bibitem[\protect\citeauthoryear{{Galloway}, {Sokoloski}  \&
  {Kenyon}}{{Galloway} et~al.}{2002}]{2002ApJ...580.1065G}
{Galloway} D.~K.,  {Sokoloski} J.~L.,   {Kenyon} S.~J.,  2002, \mn@doi [\apj]
  {10.1086/343798}, \href {http://adsabs.harvard.edu/abs/2002ApJ...580.1065G}
  {580, 1065}

\bibitem[\protect\citeauthoryear{{Gil-Pons}, {Garc{\'{\i}}a-Berro}, {Jos{\'e}},
  {Hernanz}  \& {Truran}}{{Gil-Pons} et~al.}{2003}]{2003A&A...407.1021G}
{Gil-Pons} P.,  {Garc{\'{\i}}a-Berro} E.,  {Jos{\'e}} J.,  {Hernanz} M.,
  {Truran} J.~W.,  2003, \mn@doi [\aap] {10.1051/0004-6361:20030852}, \href
  {http://esoads.eso.org/abs/2003A%26A...407.1021G} {407, 1021}

\bibitem[\protect\citeauthoryear{{Gonz{\'a}lez-Gal{\'a}n}, {Kuulkers},
  {Kretschmar}, {Larsson}, {Postnov}, {Kochetkova}  \&
  {Finger}}{{Gonz{\'a}lez-Gal{\'a}n} et~al.}{2012}]{GonzalezGalan2011}
{Gonz{\'a}lez-Gal{\'a}n} A.,  {Kuulkers} E.,  {Kretschmar} P.,  {Larsson} S.,
  {Postnov} K.,  {Kochetkova} A.,   {Finger} M.~H.,  2012, \mn@doi [\aap]
  {10.1051/0004-6361/201117893}, \href
  {http://adsabs.harvard.edu/abs/2012A%26A...537A..66G} {537, A66}

\bibitem[\protect\citeauthoryear{{Hinkle}, {Fekel}, {Joyce}, {Wood}, {Smith}
  \& {Lebzelter}}{{Hinkle} et~al.}{2006}]{2006ApJ...641..479H}
{Hinkle} K.~H.,  {Fekel} F.~C.,  {Joyce} R.~R.,  {Wood} P.~R.,  {Smith} V.~V.,
   {Lebzelter} T.,  2006, \mn@doi [\apj] {10.1086/500350}, \href
  {http://adsabs.harvard.edu/abs/2006ApJ...641..479H} {641, 479}

\bibitem[\protect\citeauthoryear{{Hinkle}, {Fekel}, {Joyce}, {Miko{\l}ajewska},
  {Galan}  \& {Lebzelter}}{{Hinkle} et~al.}{2018}]{2018arXiv181208811H}
{Hinkle} K.~H.,  {Fekel} F.~C.,  {Joyce} R.~R.,  {Miko{\l}ajewska} J.,  {Galan}
  C.,   {Lebzelter} T.,  2018, arXiv e-prints, \href
  {http://esoads.eso.org/abs/2018arXiv181208811H} {}

\bibitem[\protect\citeauthoryear{{Ho}, {Taam}, {Fryxell}, {Matsuda}  \&
  {Koide}}{{Ho} et~al.}{1989}]{1989MNRAS.238.1447H}
{Ho} C.,  {Taam} R.~E.,  {Fryxell} B.~A.,  {Matsuda} T.,   {Koide} H.,  1989,
  \mn@doi [\mnras] {10.1093/mnras/238.4.1447}, \href
  {http://adsabs.harvard.edu/abs/1989MNRAS.238.1447H} {238, 1447}

\bibitem[\protect\citeauthoryear{{Hobbs}, {Lorimer}, {Lyne}  \&
  {Kramer}}{{Hobbs} et~al.}{2005}]{2005MNRAS.360..974H}
{Hobbs} G.,  {Lorimer} D.~R.,  {Lyne} A.~G.,   {Kramer} M.,  2005, \mn@doi
  [\mnras] {10.1111/j.1365-2966.2005.09087.x}, \href
  {http://adsabs.harvard.edu/abs/2005MNRAS.360..974H} {360, 974}

\bibitem[\protect\citeauthoryear{{Hubrig}, {Sidoli}, {Postnov}, {Sch{\"o}ller},
  {Kholtygin}, {J{\"a}rvinen}  \& {Steinbrunner}}{{Hubrig}
  et~al.}{2018}]{2018MNRAS.474L..27H}
{Hubrig} S.,  {Sidoli} L.,  {Postnov} K.,  {Sch{\"o}ller} M.,  {Kholtygin}
  A.~F.,  {J{\"a}rvinen} S.~P.,   {Steinbrunner} P.,  2018, \mn@doi [\mnras]
  {10.1093/mnrasl/slx187}, \href
  {http://adsabs.harvard.edu/abs/2018MNRAS.474L..27H} {474, L27}

\bibitem[\protect\citeauthoryear{{Hunt}}{{Hunt}}{1971}]{1971MNRAS.154..141H}
{Hunt} R.,  1971, \mn@doi [\mnras] {10.1093/mnras/154.2.141}, \href
  {http://adsabs.harvard.edu/abs/1971MNRAS.154..141H} {154, 141}

\bibitem[\protect\citeauthoryear{{Hurley}, {Pols}  \& {Tout}}{{Hurley}
  et~al.}{2000}]{2000MNRAS.315..543H}
{Hurley} J.~R.,  {Pols} O.~R.,   {Tout} C.~A.,  2000, \mn@doi [\mnras]
  {10.1046/j.1365-8711.2000.03426.x}, \href
  {http://adsabs.harvard.edu/abs/2000MNRAS.315..543H} {315, 543}

\bibitem[\protect\citeauthoryear{{Hurley}, {Tout}  \& {Pols}}{{Hurley}
  et~al.}{2002}]{2002MNRAS.329..897H}
{Hurley} J.~R.,  {Tout} C.~A.,   {Pols} O.~R.,  2002, \mn@doi [\mnras]
  {10.1046/j.1365-8711.2002.05038.x}, \href
  {http://adsabs.harvard.edu/abs/2002MNRAS.329..897H} {329, 897}

\bibitem[\protect\citeauthoryear{{Hurley}, {Tout}, {Wickramasinghe}, {Ferrario}
   \& {Kiel}}{{Hurley} et~al.}{2010}]{2010MNRAS.402.1437H}
{Hurley} J.~R.,  {Tout} C.~A.,  {Wickramasinghe} D.~T.,  {Ferrario} L.,
  {Kiel} P.~D.,  2010, \mn@doi [\mnras] {10.1111/j.1365-2966.2009.15988.x},
  \href {http://esoads.eso.org/abs/2010MNRAS.402.1437H} {402, 1437}

\bibitem[\protect\citeauthoryear{{Hynes} et~al.,}{{Hynes}
  et~al.}{2014}]{2014ApJ...780...11H}
{Hynes} R.~I.,  et~al., 2014, \mn@doi [\apj] {10.1088/0004-637X/780/1/11},
  \href {http://adsabs.harvard.edu/abs/2014ApJ...780...11H} {780, 11}

\bibitem[\protect\citeauthoryear{{Hynes} et~al.,}{{Hynes}
  et~al.}{2017}]{2017AAS...23031704H}
{Hynes} R.~I.,  et~al., 2017, in AAS Meeting Abstracts \#230. p. 317.04

\bibitem[\protect\citeauthoryear{{Iben} \& {Tutukov}}{{Iben} \&
  {Tutukov}}{1996}]{1996ApJS..105..145I}
{Iben} Jr. I.,  {Tutukov} A.~V.,  1996, \mn@doi [\apjs] {10.1086/192310}, \href
  {http://esoads.eso.org/abs/1996ApJS..105..145I} {105, 145}

\bibitem[\protect\citeauthoryear{{I{\l}kiewicz}, {Miko{\l}ajewska}  \&
  {Monard}}{{I{\l}kiewicz} et~al.}{2017}]{2017A&A...601A.105I}
{I{\l}kiewicz} K.,  {Miko{\l}ajewska} J.,   {Monard} B.,  2017, \mn@doi [\aap]
  {10.1051/0004-6361/201630021}, \href
  {http://esoads.eso.org/abs/2017A%26A...601A.105I} {601, A105}

\bibitem[\protect\citeauthoryear{{Illarionov} \& {Sunyaev}}{{Illarionov} \&
  {Sunyaev}}{1975}]{1975A&A....39..185I}
{Illarionov} A.~F.,  {Sunyaev} R.~A.,  1975, \aap, \href
  {http://adsabs.harvard.edu/abs/1975A%26A....39..185I} {39, 185}

\bibitem[\protect\citeauthoryear{{Jones}, {R{\"o}pke}, {Pakmor}, {Seitenzahl},
  {Ohlmann}  \& {Edelmann}}{{Jones} et~al.}{2016}]{2016A&A...593A..72J}
{Jones} S.,  {R{\"o}pke} F.~K.,  {Pakmor} R.,  {Seitenzahl} I.~R.,  {Ohlmann}
  S.~T.,   {Edelmann} P.~V.~F.,  2016, \mn@doi [\aap]
  {10.1051/0004-6361/201628321}, \href
  {http://esoads.eso.org/abs/2016A%26A...593A..72J} {593, A72}

\bibitem[\protect\citeauthoryear{{Jose} \& {Hernanz}}{{Jose} \&
  {Hernanz}}{1998}]{1998ApJ...494..680J}
{Jose} J.,  {Hernanz} M.,  1998, \mn@doi [\apj] {10.1086/305244}, \href
  {http://adsabs.harvard.edu/abs/1998ApJ...494..680J} {494, 680}

\bibitem[\protect\citeauthoryear{{Kaplan}, {Levine}, {Chakrabarty}, {Morgan},
  {Erb}, {Gaensler}, {Moon}  \& {Cameron}}{{Kaplan}
  et~al.}{2007}]{2007ApJ...661..437K}
{Kaplan} D.~L.,  {Levine} A.~M.,  {Chakrabarty} D.,  {Morgan} E.~H.,  {Erb}
  D.~K.,  {Gaensler} B.~M.,  {Moon} D.-S.,   {Cameron} P.~B.,  2007, \mn@doi
  [\apj] {10.1086/513712}, \href
  {http://adsabs.harvard.edu/abs/2007ApJ...661..437K} {661, 437}

\bibitem[\protect\citeauthoryear{{Kenyon} \& {Webbink}}{{Kenyon} \&
  {Webbink}}{1984}]{1984ApJ...279..252K}
{Kenyon} S.~J.,  {Webbink} R.~F.,  1984, \mn@doi [\apj] {10.1086/161888}, \href
  {http://esoads.eso.org/abs/1984ApJ...279..252K} {279, 252}

\bibitem[\protect\citeauthoryear{{Kudritzki} \& {Reimers}}{{Kudritzki} \&
  {Reimers}}{1978}]{1978A&A....70..227K}
{Kudritzki} R.~P.,  {Reimers} D.,  1978, \aap, \href
  {http://adsabs.harvard.edu/abs/1978A%26A....70..227K} {70, 227}

\bibitem[\protect\citeauthoryear{{Kuranov} \& {Postnov}}{{Kuranov} \&
  {Postnov}}{2015}]{2015AstL...41..114K}
{Kuranov} A.~G.,  {Postnov} K.~A.,  2015, \mn@doi [Astronomy Letters]
  {10.1134/S1063773715040064}, \href
  {http://adsabs.harvard.edu/abs/2015AstL...41..114K} {41, 114}

\bibitem[\protect\citeauthoryear{{Lipunov}, {Postnov}, {Prokhorov}  \&
  {Bogomazov}}{{Lipunov} et~al.}{2009}]{2009ARep...53..915L}
{Lipunov} V.~M.,  {Postnov} K.~A.,  {Prokhorov} M.~E.,   {Bogomazov} A.~I.,
  2009, \mn@doi [Astronomy Reports] {10.1134/S1063772909100047}, \href
  {http://adsabs.harvard.edu/abs/2009ARep...53..915L} {53, 915}

\bibitem[\protect\citeauthoryear{{Liu}, {Stancliffe}, {Abate}  \&
  {Matrozis}}{{Liu} et~al.}{2017}]{2017ApJ...846..117L}
{Liu} Z.-W.,  {Stancliffe} R.~J.,  {Abate} C.,   {Matrozis} E.,  2017, \mn@doi
  [\apj] {10.3847/1538-4357/aa8622}, \href
  {http://adsabs.harvard.edu/abs/2017ApJ...846..117L} {846, 117}

\bibitem[\protect\citeauthoryear{{Loveridge}, {van der Sluys}  \&
  {Kalogera}}{{Loveridge} et~al.}{2011}]{2011ApJ...743...49L}
{Loveridge} A.~J.,  {van der Sluys} M.~V.,   {Kalogera} V.,  2011, \mn@doi
  [\apj] {10.1088/0004-637X/743/1/49}, \href
  {http://adsabs.harvard.edu/abs/2011ApJ...743...49L} {743, 49}

\bibitem[\protect\citeauthoryear{{L{\"u}}, {Yungelson}  \& {Han}}{{L{\"u}}
  et~al.}{2006}]{2006MNRAS.372.1389L}
{L{\"u}} G.,  {Yungelson} L.,   {Han} Z.,  2006, \mn@doi [\mnras]
  {10.1111/j.1365-2966.2006.10947.x}, \href
  {http://adsabs.harvard.edu/abs/2006MNRAS.372.1389L} {372, 1389}

\bibitem[\protect\citeauthoryear{{L{\"u}}, {Zhu}, {Postnov}, {Yungelson},
  {Kuranov}  \& {Wang}}{{L{\"u}} et~al.}{2012}]{2012MNRAS.424.2265L}
{L{\"u}} G.-L.,  {Zhu} C.-H.,  {Postnov} K.~A.,  {Yungelson} L.~R.,  {Kuranov}
  A.~G.,   {Wang} N.,  2012, \mn@doi [\mnras]
  {10.1111/j.1365-2966.2012.21395.x}, \href
  {http://esoads.eso.org/abs/2012MNRAS.424.2265L} {424, 2265}

\bibitem[\protect\citeauthoryear{{Lutovinov}, {Revnivtsev}, {Gilfanov},
  {Shtykovskiy}, {Molkov}  \& {Sunyaev}}{{Lutovinov}
  et~al.}{2005}]{2005A&A...444..821L}
{Lutovinov} A.,  {Revnivtsev} M.,  {Gilfanov} M.,  {Shtykovskiy} P.,  {Molkov}
  S.,   {Sunyaev} R.,  2005, \mn@doi [\aap] {10.1051/0004-6361:20042392}, \href
  {http://adsabs.harvard.edu/abs/2005A%26A...444..821L} {444, 821}

\bibitem[\protect\citeauthoryear{{Majczyna}, {Madej}, {Nale{\.z}yty},
  {R{\'o}{\.z}a{\'n}ska}  \& {Udalski}}{{Majczyna}
  et~al.}{2016}]{2016pas..conf..133M}
{Majczyna} A.,  {Madej} J.,  {Nale{\.z}yty} M.,  {R{\'o}{\.z}a{\'n}ska} A.,
  {Udalski} A.,  2016, in {R{\'o}{\.z}a{\'n}ska} A.,  {Bejger} M.,  eds, 37th
  Meeting of the Polish Astronomical Society. pp 133--136

\bibitem[\protect\citeauthoryear{{Masetti} et~al.,}{{Masetti}
  et~al.}{2002}]{2002A&A...382..104M}
{Masetti} N.,  et~al., 2002, \mn@doi [\aap] {10.1051/0004-6361:20011543}, \href
  {http://adsabs.harvard.edu/abs/2002A%26A...382..104M} {382, 104}

\bibitem[\protect\citeauthoryear{{Masetti}, {Orlandini}, {Palazzi}, {Amati}  \&
  {Frontera}}{{Masetti} et~al.}{2006}]{2006A&A...453..295M}
{Masetti} N.,  {Orlandini} M.,  {Palazzi} E.,  {Amati} L.,   {Frontera} F.,
  2006, \mn@doi [\aap] {10.1051/0004-6361:20065025}, \href
  {http://adsabs.harvard.edu/abs/2006A%26A...453..295M} {453, 295}

\bibitem[\protect\citeauthoryear{{Masetti} et~al.,}{{Masetti}
  et~al.}{2007}]{2007A&A...470..331M}
{Masetti} N.,  et~al., 2007, \mn@doi [\aap] {10.1051/0004-6361:20077509}, \href
  {http://adsabs.harvard.edu/abs/2007A%26A...470..331M} {470, 331}

\bibitem[\protect\citeauthoryear{{Masetti}, {Munari}, {Henden}, {Page},
  {Osborne}  \& {Starrfield}}{{Masetti} et~al.}{2011}]{2011A&A...534A..89M}
{Masetti} N.,  {Munari} U.,  {Henden} A.~A.,  {Page} K.~L.,  {Osborne} J.~P.,
  {Starrfield} S.,  2011, \mn@doi [\aap] {10.1051/0004-6361/201117260}, \href
  {http://esoads.eso.org/abs/2011A%26A...534A..89M} {534, A89}

\bibitem[\protect\citeauthoryear{{Masetti} et~al.,}{{Masetti}
  et~al.}{2012}]{2012A&A...538A.123M}
{Masetti} N.,  et~al., 2012, \mn@doi [\aap] {10.1051/0004-6361/201118559},
  \href {http://adsabs.harvard.edu/abs/2012A%26A...538A.123M} {538, A123}

\bibitem[\protect\citeauthoryear{{Miko{\l}ajewska}}{{Miko{\l}ajewska}}{2012}]{2012BaltA..21....5M}
{Miko{\l}ajewska} J.,  2012, Baltic Astronomy, \href
  {http://esoads.eso.org/abs/2012BaltA..21....5M} {21, 5}

\bibitem[\protect\citeauthoryear{{Miyaji}, {Nomoto}, {Yokoi}  \&
  {Sugimoto}}{{Miyaji} et~al.}{1980}]{1980PASJ...32..303M}
{Miyaji} S.,  {Nomoto} K.,  {Yokoi} K.,   {Sugimoto} D.,  1980, \pasj, \href
  {http://esoads.eso.org/abs/1980PASJ...32..303M} {32, 303}

\bibitem[\protect\citeauthoryear{{Nelemans} \& {Tout}}{{Nelemans} \&
  {Tout}}{2005}]{2005MNRAS.356..753N}
{Nelemans} G.,  {Tout} C.~A.,  2005, \mn@doi [\mnras]
  {10.1111/j.1365-2966.2004.08496.x}, \href
  {http://adsabs.harvard.edu/abs/2005MNRAS.356..753N} {356, 753}

\bibitem[\protect\citeauthoryear{{Nelemans}, {Verbunt}, {Yungelson}  \&
  {Portegies Zwart}}{{Nelemans} et~al.}{2000}]{2000A&A...360.1011N}
{Nelemans} G.,  {Verbunt} F.,  {Yungelson} L.~R.,   {Portegies Zwart} S.~F.,
  2000, \aap, \href {http://adsabs.harvard.edu/abs/2000A%26A...360.1011N} {360,
  1011}

\bibitem[\protect\citeauthoryear{{Ohlmann}, {R{\"o}pke}, {Pakmor}  \&
  {Springel}}{{Ohlmann} et~al.}{2016}]{2016ApJ...816L...9O}
{Ohlmann} S.~T.,  {R{\"o}pke} F.~K.,  {Pakmor} R.,   {Springel} V.,  2016,
  \mn@doi [\apjl] {10.3847/2041-8205/816/1/L9}, \href
  {http://esoads.eso.org/abs/2016ApJ...816L...9O} {816, L9}

\bibitem[\protect\citeauthoryear{{Patel} et~al.,}{{Patel}
  et~al.}{2004}]{2004ApJ...602L..45P}
{Patel} S.~K.,  et~al., 2004, \mn@doi [\apjl] {10.1086/382210}, \href
  {http://adsabs.harvard.edu/abs/2004ApJ...602L..45P} {602, L45}

\bibitem[\protect\citeauthoryear{{Patel} et~al.,}{{Patel}
  et~al.}{2007}]{2007ApJ...657..994P}
{Patel} S.~K.,  et~al., 2007, \mn@doi [\apj] {10.1086/510374}, \href
  {http://adsabs.harvard.edu/abs/2007ApJ...657..994P} {657, 994}

\bibitem[\protect\citeauthoryear{{Pereira}, {Braga}  \& {Jablonski}}{{Pereira}
  et~al.}{1999}]{1999ApJ...526L.105P}
{Pereira} M.~G.,  {Braga} J.,   {Jablonski} F.,  1999, \mn@doi [\apjl]
  {10.1086/312368}, \href {http://adsabs.harvard.edu/abs/1999ApJ...526L.105P}
  {526, L105}

\bibitem[\protect\citeauthoryear{{Piro} \& {Thompson}}{{Piro} \&
  {Thompson}}{2014}]{2014ApJ...794...28P}
{Piro} A.~L.,  {Thompson} T.~A.,  2014, \mn@doi [\apj]
  {10.1088/0004-637X/794/1/28}, \href
  {http://esoads.eso.org/abs/2014ApJ...794...28P} {794, 28}

\bibitem[\protect\citeauthoryear{{Postnov} \& {Yungelson}}{{Postnov} \&
  {Yungelson}}{2014}]{2014LRR....17....3P}
{Postnov} K.~A.,  {Yungelson} L.~R.,  2014, \mn@doi [Living Reviews in
  Relativity] {10.12942/lrr-2014-3}, \href
  {http://adsabs.harvard.edu/abs/2014LRR....17....3P} {17, 3}

\bibitem[\protect\citeauthoryear{{Postnov}, {Kuranov}, {Kolesnikov}, {Popov}
  \& {Porayko}}{{Postnov} et~al.}{2016}]{2016MNRAS.463.1642P}
{Postnov} K.~A.,  {Kuranov} A.~G.,  {Kolesnikov} D.~A.,  {Popov} S.~B.,
  {Porayko} N.~K.,  2016, \mn@doi [\mnras] {10.1093/mnras/stw2080}, \href
  {http://adsabs.harvard.edu/abs/2016MNRAS.463.1642P} {463, 1642}

\bibitem[\protect\citeauthoryear{{Postnov}, {Kuranov}  \&
  {Yungelson}}{{Postnov} et~al.}{2018}]{2018arXiv181102842P}
{Postnov} K.,  {Kuranov} A.,   {Yungelson} L.,  2018, preprint, \href
  {http://adsabs.harvard.edu/abs/2018arXiv181102842P} {} (\mn@eprint {arXiv}
  {1811.02842})

\bibitem[\protect\citeauthoryear{{Predehl} et~al.,}{{Predehl}
  et~al.}{2016}]{2016SPIE.9905E..1KP}
{Predehl} P.,  et~al., 2016, in Space Telescopes and Instrumentation 2016:
  Ultraviolet to Gamma Ray. p. 99051K, \mn@doi{10.1117/12.2235092}

\bibitem[\protect\citeauthoryear{{Predehl} et~al.,}{{Predehl}
  et~al.}{2018}]{2018SPIE10699E..5HP}
{Predehl} P.,  et~al., 2018, in Space Telescopes and Instrumentation 2018:
  Ultraviolet to Gamma Ray. p. 106995H, \mn@doi{10.1117/12.2315139}

\bibitem[\protect\citeauthoryear{{Qiu}, {Zhou}, {Yu}, {Li}  \& {Xu}}{{Qiu}
  et~al.}{2017}]{2017ApJ...847...44Q}
{Qiu} H.,  {Zhou} P.,  {Yu} W.,  {Li} X.,   {Xu} X.,  2017, \mn@doi [\apj]
  {10.3847/1538-4357/aa8728}, \href
  {http://adsabs.harvard.edu/abs/2017ApJ...847...44Q} {847, 44}

\bibitem[\protect\citeauthoryear{{Ricker}, {Timmes}, {Taam}  \&
  {Webbink}}{{Ricker} et~al.}{2018}]{2018arXiv181103656R}
{Ricker} P.~M.,  {Timmes} F.~X.,  {Taam} R.~E.,   {Webbink} R.~F.,  2018,
  preprint, \href {http://adsabs.harvard.edu/abs/2018arXiv181103656R} {}
  (\mn@eprint {arXiv} {1811.03656})

\bibitem[\protect\citeauthoryear{{Sana} et~al.,}{{Sana}
  et~al.}{2012}]{2012Sci...337..444S}
{Sana} H.,  et~al., 2012, \mn@doi [Science] {10.1126/science.1223344}, \href
  {http://adsabs.harvard.edu/abs/2012Sci...337..444S} {337, 444}

\bibitem[\protect\citeauthoryear{{Schwab} \& {Akira Rocha}}{{Schwab} \& {Akira
  Rocha}}{2018}]{2018arXiv181013002S}
{Schwab} J.,  {Akira Rocha} K.,  2018, arXiv e-prints, \href
  {http://esoads.eso.org/abs/2018arXiv181013002S} {}

\bibitem[\protect\citeauthoryear{{Schwab}, {Quataert}  \& {Bildsten}}{{Schwab}
  et~al.}{2015}]{2015MNRAS.453.1910S}
{Schwab} J.,  {Quataert} E.,   {Bildsten} L.,  2015, \mn@doi [\mnras]
  {10.1093/mnras/stv1804}, \href
  {http://esoads.eso.org/abs/2015MNRAS.453.1910S} {453, 1910}

\bibitem[\protect\citeauthoryear{{Shakura} \& {Postnov}}{{Shakura} \&
  {Postnov}}{2017}]{2017arXiv170203393S}
{Shakura} N.,  {Postnov} K.,  2017, preprint, \href
  {http://adsabs.harvard.edu/abs/2017arXiv170203393S} {} (\mn@eprint {arXiv}
  {1702.03393})

\bibitem[\protect\citeauthoryear{{Shakura}, {Postnov}, {Kochetkova}  \&
  {Hjalmarsdotter}}{{Shakura} et~al.}{2012}]{2012MNRAS.420..216S}
{Shakura} N.,  {Postnov} K.,  {Kochetkova} A.,   {Hjalmarsdotter} L.,  2012,
  \mn@doi [\mnras] {10.1111/j.1365-2966.2011.20026.x}, \href
  {http://esoads.eso.org/abs/2012MNRAS.420..216S} {420, 216}

\bibitem[\protect\citeauthoryear{{Shakura}, {Postnov}, {Sidoli}  \&
  {Paizis}}{{Shakura} et~al.}{2014}]{2014MNRAS.442.2325S}
{Shakura} N.,  {Postnov} K.,  {Sidoli} L.,   {Paizis} A.,  2014, \mn@doi
  [\mnras] {10.1093/mnras/stu1027}, \href
  {http://adsabs.harvard.edu/abs/2014MNRAS.442.2325S} {442, 2325}

\bibitem[\protect\citeauthoryear{{Shakura}, {Postnov}, {Kochetkova}  \&
  {Hjalmarsdotter}}{{Shakura} et~al.}{2018}]{2018ASSL..454..331S}
{Shakura} N.,  {Postnov} K.,  {Kochetkova} A.,   {Hjalmarsdotter} L.,  2018, in
  {Shakura} N.,  ed.,  Astrophysics and Space Science Library Vol. 454,
  Astrophysics and Space Science Library. p.~331,
  \mn@doi{10.1007/978-3-319-93009-1_7}

\bibitem[\protect\citeauthoryear{{Siess} \& {Lebreuilly}}{{Siess} \&
  {Lebreuilly}}{2018}]{2018A&A...614A..99S}
{Siess} L.,  {Lebreuilly} U.,  2018, \mn@doi [\aap]
  {10.1051/0004-6361/201732502}, \href
  {http://adsabs.harvard.edu/abs/2018A%26A...614A..99S} {614, A99}

\bibitem[\protect\citeauthoryear{{Smith}, {Markwardt}, {Swank}  \&
  {Negueruela}}{{Smith} et~al.}{2012}]{2012MNRAS.422.2661S}
{Smith} D.~M.,  {Markwardt} C.~B.,  {Swank} J.~H.,   {Negueruela} I.,  2012,
  \mn@doi [\mnras] {10.1111/j.1365-2966.2012.20836.x}, \href
  {http://esoads.eso.org/abs/2012MNRAS.422.2661S} {422, 2661}

\bibitem[\protect\citeauthoryear{{Vink}}{{Vink}}{2017}]{2017A&A...607L...8V}
{Vink} J.~S.,  2017, \mn@doi [\aap] {10.1051/0004-6361/201731902}, \href
  {http://adsabs.harvard.edu/abs/2017A%26A...607L...8V} {607, L8}

\bibitem[\protect\citeauthoryear{{Vink}}{{Vink}}{2018}]{2018A&A...619A..54V}
{Vink} J.~S.,  2018, \mn@doi [\aap] {10.1051/0004-6361/201833352}, \href
  {http://esoads.eso.org/abs/2018A%26A...619A..54V} {619, A54}

\bibitem[\protect\citeauthoryear{Webbink}{Webbink}{1984}]{web84}
Webbink R.~F.,  1984, \apj, 277, 355

\bibitem[\protect\citeauthoryear{{Wilson}}{{Wilson}}{1971}]{1971ApJ...163..209W}
{Wilson} J.~R.,  1971, \mn@doi [\apj] {10.1086/150759}, \href
  {http://esoads.eso.org/abs/1971ApJ...163..209W} {163, 209}

\bibitem[\protect\citeauthoryear{{Yu} \& {Jeffery}}{{Yu} \&
  {Jeffery}}{2010}]{2010A&A...521A..85Y}
{Yu} S.,  {Jeffery} C.~S.,  2010, \mn@doi [\aap] {10.1051/0004-6361/201014827},
  \href {http://adsabs.harvard.edu/abs/2010A%26A...521A..85Y} {521, A85}

\bibitem[\protect\citeauthoryear{{Yungelson} \& {Kuranov}}{{Yungelson} \&
  {Kuranov}}{2017}]{2017MNRAS.464.1607Y}
{Yungelson} L.~R.,  {Kuranov} A.~G.,  2017, \mn@doi [\mnras]
  {10.1093/mnras/stw2432}, \href
  {http://adsabs.harvard.edu/abs/2017MNRAS.464.1607Y} {464, 1607}

\bibitem[\protect\citeauthoryear{{de Val-Borro}, {Karovska}, {Sasselov}  \&
  {Stone}}{{de Val-Borro} et~al.}{2017}]{2017MNRAS.468.3408D}
{de Val-Borro} M.,  {Karovska} M.,  {Sasselov} D.~D.,   {Stone} J.~M.,  2017,
  \mn@doi [\mnras] {10.1093/mnras/stx684}, \href
  {http://adsabs.harvard.edu/abs/2017MNRAS.468.3408D} {468, 3408}

\makeatother
\end{thebibliography}



\appendix

\section{Effect of the orbital eccentricity on the NS equilibrium period at the settling accretion stage}

\begin{figure*}
\includegraphics[width=0.49\textwidth]{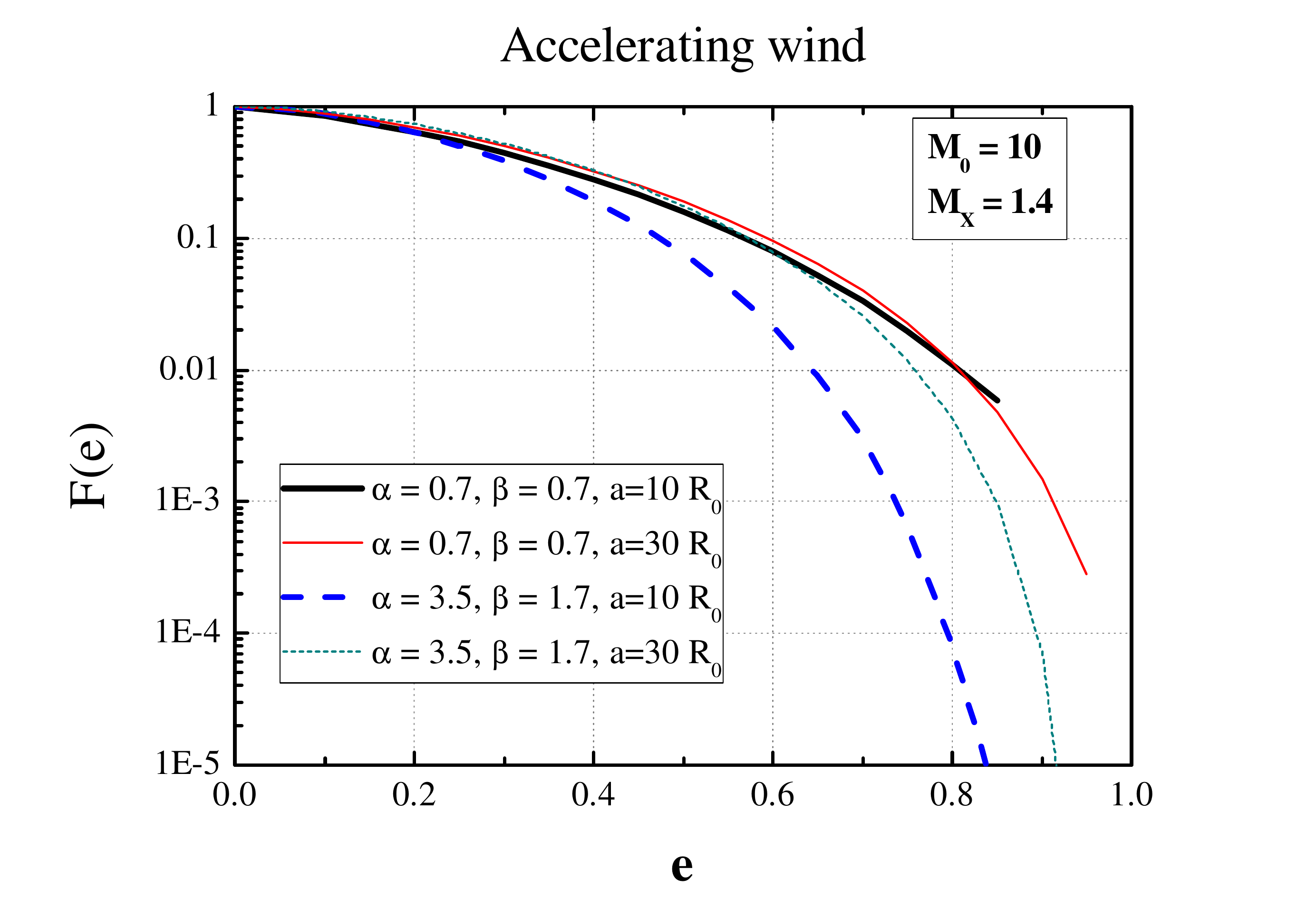}
\includegraphics[width=0.49\textwidth]{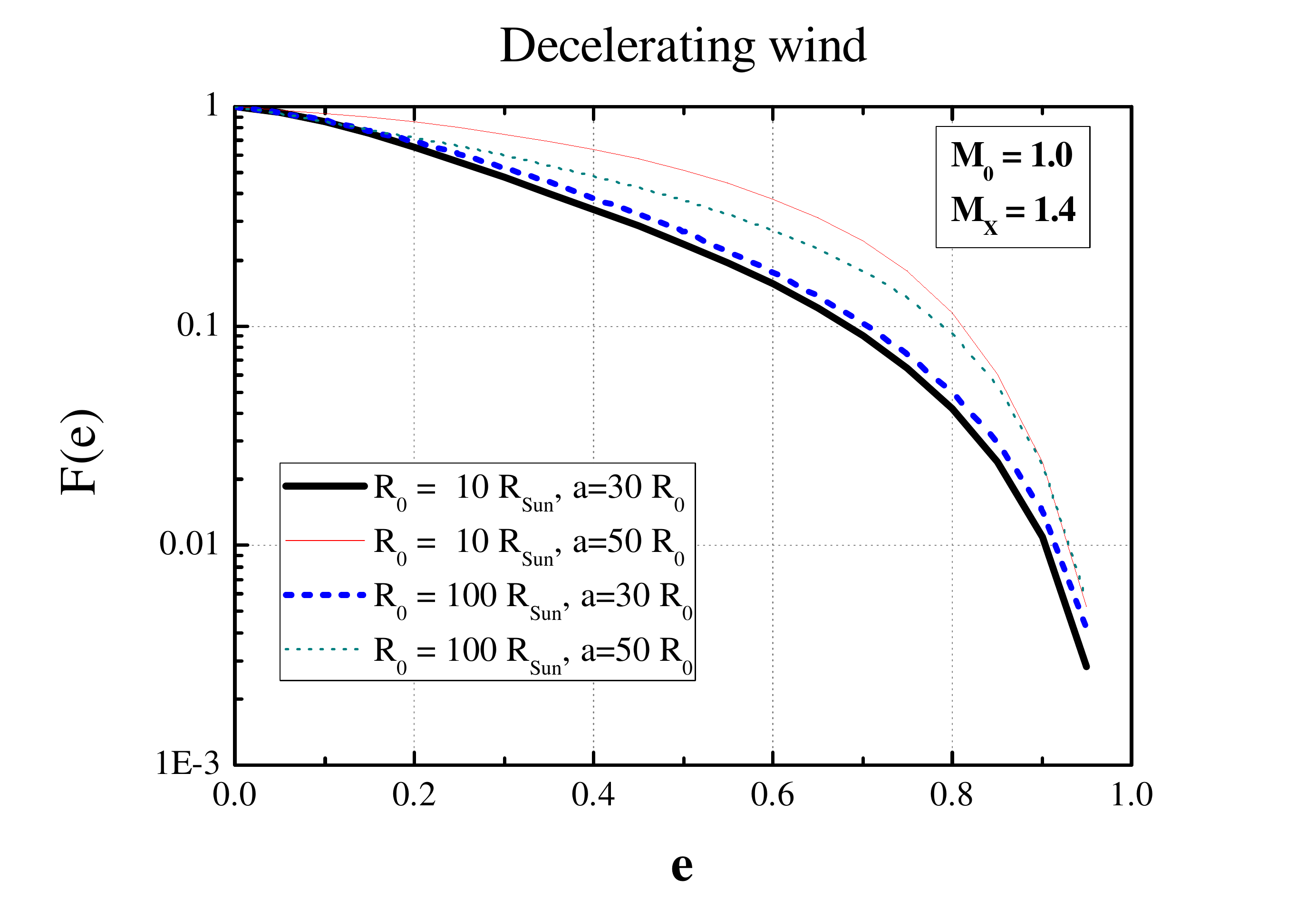}
\caption{Reduction factor $F(e)$ as a function of the orbital eccentricity $e$ for several values of 
$a/R_\mathrm{o}$, $M_\mathrm{o}/M$. Left: the case of radiatively-driven accelerating winds
 [Eq.~\ref{a:frearly}].
Right: the case of decelerating winds from evolved red giants [Eq.~\ref{a:frdec}].
}
\label{f:ax}
\end{figure*}

Here we describe the effect of orbital eccentricity on the value of the equilibrium NS period at the settling accretion stage. Because of the orbital eccentricity, the spin-up and spin-down torques applied to the NS change during the orbital motion and should be averaged over the orbital period. In the Keplerian two-body problem, the separation between  the barycentres of the binary components is 
\beq{a:r}
r=\frac{p}{1+e\cos\theta},
\eeq
where $p=a(1-e^2)$ is the orbital semilatus rectum, $a$ is the semi-major axis of the orbit, $e$ is the orbital eccentricity, $\theta$ is the true anomaly. The stellar wind velocity is assumed to be spherically symmetric and centered on the optical star barycentre. 
If the radius of the optical star is $R_\mathrm{o}$, the stellar wind velocity as a function of distance from the star, normalized to the escape velocity at the optical star's photosphere, $v_\mathrm{esc}=\sqrt{2GM_\mathrm{o}/R_\mathrm{o}}$, can be written as 
\beq{a:vw}
v_\mathrm{w}(r)=v_\mathrm{esc}f(r). 
\eeq
For example, for radiative-driven accelerating winds of early type stars
\beq{a:frearly}
f(r) = \alpha \left(1-\frac{R_\mathrm{o}}{r}\right)^\beta\,,
\eeq
where $\alpha=0.7...3.8$ and $\beta=0.7...2.0$ are obtained from recent numerical simulations  \citep{2018A&A...619A..54V}.

For decelerating winds from evolved red giants -- optical components of SyXBs, we follow the prescription given in \cite{2006MNRAS.372.1389L},
which can be  conveniently written as a continuous function
\beq{a:frdec}
f(r)=\left(1-\frac{5 \kms}{v_\mathrm{esc}}\right)\exp{\left(-\left[\frac{1}{5}\left(\frac{r}{R_\mathrm{o}}-1\right)\right]^2\right)}+ \frac{5 \kms}{v_\mathrm{esc}}\,.
\eeq

In the case of an eccentric orbit, the specific angular momentum of captured wind matter is determined by the tangent component of the relative orbital velocity of the NS, 
\beq{a:vn}
v_n=\sqrt{\frac{GM}{p}}(1+e\cos\theta)\,,
\eeq
while the gravitational capture Bondi radius $R_\mathrm{B}$ depends on the module of the sum of the orbital velocity vector 
$\bf{v}=\bf{v}_n+\bf{v}_r$ and the wind velocity vector $\bf{v}_\mathrm{w}(r)$ that has the radial component only:
\beq{a:v}
v(r)=\sqrt{(v_\mathrm{w}(r)-v_r)^2+v_n^2}\,.
\eeq
Here $v_r$ is the radial orbital velocity component:
\beq{a:vr}
v_r=\sqrt{\frac{GM}{p}}(e\sin\theta)\,.
\eeq

In the case of an eccentric orbit, the spin-up torque in 
\Eq{e:spinev} takes the form:
\beq{a:sd}
K_{sd}(r)=Z\dot M_\mathrm{x} v_n R_\mathrm{B} (R_\mathrm{B}/r)
\eeq
and therefore (neglecting  the term $z/Z\ll 1$) the equilibrium NS period averaged over the orbit can be determined from the condition $\langle dI\omega^*/dt\rangle =0$:
\beq{a:peq}
P_\mathrm{eq}^{e}=2\piup\frac{\left\langle 
\dot M_\mathrm{B} R_\mathrm{A}^2
\right\rangle}{\left\langle \dot M_\mathrm{B} v_nR_\mathrm{B}^2/r\right\rangle}\,.
\eeq
Here $\langle \ldots\rangle=2\int_0^{\piup}\ldots d\theta$.

If we take into account mass conservation of spherically symmetric stellar wind of an optical star losing mass at the rate 
$\dot M_\mathrm{o}$, i.e., 
\begin{equation}
\label{a:RB}
\dot{M}_\mathrm{B}(r)=\frac{1}{4}\dot{M}_\mathrm{o}\left(\frac{v(r)}{v_\mathrm{w}(r)}\right)
\left(\frac{R_\mathrm{B}(r)}{r}\right)^2\,,
\end{equation} 
and keep in mind that the Alfv\'en radius $R_\mathrm{A}\sim \dot{M}_\mathrm{B}^{-2/7}$, 
we arrive at the ratio of the NS equilibrium period in an eccentric orbit to that in the circular orbit 
($e=0$, $P_\mathrm{eq}^0=P_\mathrm{B}(R_\mathrm{A}/R_\mathrm{B})^2)$:  
\beq{a:x}
F(e)=\mathlarger{
\frac{\sqrt{\frac{GM}{a}}}{v_0}
\frac{\left\langle\myfrac{1+e\cos\theta}{1-e^2}^{6/7} \myfrac{v}{v_\mathrm{w}}^{3/7}\myfrac{v_0}{v}^{12/7}
\right\rangle \myfrac{v_0}{v_{w,0}}^{4/7}}
{\left\langle 
\myfrac{1+e\cos\theta}{1-e^2}^{3} \myfrac{v_n}{v_\mathrm{w}}\myfrac{v_0}{v}^{7}
\right\rangle} \le1}.
\eeq
Here the relative NS orbital velocity in the circular orbit $r=a$ is
\beq{a:v0}
v_0=\sqrt{\frac{GM}{a}}\left[\frac{2M_\mathrm{o}}{M}\frac{a}{R_\mathrm{o}}f(a)^2+1\right]^{1/2}
\eeq
and $v_{w,0}=\sqrt{\frac{2GM_\mathrm{o}}{R_\mathrm{o}}}f(a)$ is the wind velocity for the circular orbit. 

In Fig. \ref{f:ax} we plot the reduction factor $F(e)$ as a function of the orbital eccentricity $e$ for several values of $a/R_\mathrm{o}$, $M_\mathrm{o}/M$ and different stellar wind velocity laws 
[\Eq{a:frearly} and \Eq{a:frdec} for accelerating and decelerating stellar winds, respectively]. It is seen that in eccentric binaries the NS equilibrium period during the quasi-spherical settling accretion stage can be reduced by a factor of 10-100. This reduction is generally stronger for accelerating radiation-driven winds (left panel of Fig. \ref{f:ax}) and is important for high-mass X-ray binaries  \citep{2018arXiv181102842P}. For decelerating winds from late-type giants (this paper) this factor is less than 10 for  orbital eccentricities below about 0.6.

\bsp	
\label{lastpage}
\end{document}